
%
%
%
%
%
\newcount\driver \newcount\mgnf
\driver=1 \mgnf=1

\ifnum\driver=0 \special{ps: plotfile ini.pst global} \fi
\ifnum\driver=1  \fi

\newdimen\xshift \newdimen\xwidth \newdimen\yshift
\def\ins#1#2#3{\vbox to0pt{\kern-#2 \hbox{\kern#1 #3}\vss}\nointerlineskip}
\def\insertplot#1#2#3#4#5{\par%
\xwidth=#1 \xshift=\hsize \advance\xshift by-\xwidth \divide\xshift by 2%
\yshift=#2 \divide\yshift by 2%
\line{\hskip\xshift \vbox to #2{\vfil%
\ifnum\driver=0 #3
\special{ps: plotfile #4.ps} 
\fi
\ifnum\driver=1 #3 \includegraphics{#4.ps}\fi
\ifnum\driver=2#3 \special{
\ifnum\mgnf=0 #4.ps 1. 1. scale \fi
\ifnum\mgnf=1 #4.ps 1.2 1.2 scale}\fi
\fi
}\hfill\raise\yshift\hbox{#5}}}

\ifnum\mgnf=0
\magnification=\magstep0
\voffset=2.5truecm\hoffset=2.1truecm\hsize=11.7truecm \vsize=18.9truecm
\baselineskip=12pt  
\parindent=12pt
\lineskip=4pt\lineskiplimit=0.1pt      \parskip=0.1pt plus1pt
\fi
\ifnum\mgnf=1
\magnification=\magstep1\hoffset=0.cm
\voffset=-1.5truecm\hoffset=-.5truecm\hsize=16.5truecm \vsize=24.truecm
\baselineskip=14pt  
\parindent=12pt
\lineskip=4pt\lineskiplimit=0.1pt      \parskip=0.1pt plus1pt
\fi

\let\a=\alpha \let\b=\beta  \let\c=\chi \let\d=\delta  \let\e=\varepsilon
\let\f=\varphi \let\g=\gamma       \let\l=\lambda
\let\m=\mu   \let\n=\nu       \let\p=\pi  
\let\r=\rho  \let\s=\sigma    \let\th=\vartheta
 \let\x=\xi \let\z=\zeta
\let\D=\Delta     \let\L=\Lambda 
     \let\Si=\Sigma

\def\data{\number\day/\ifcase\month\or gennaio \or febbraio \or marzo \or
aprile \or maggio \or giugno \or luglio \or agosto \or settembre
\or ottobre \or novembre \or dicembre \fi/\number\year;\,\the\time}

\setbox200\hbox{$\scriptscriptstyle \data $}

\newcount\pgn \pgn=1
\def\foglio{\number\numsec:\number\pgn
\global\advance\pgn by 1}
\def\foglioa{A\number\numsec:\number\pgn
\global\advance\pgn by 1}

\global\newcount\numsec\global\newcount\numfor
\global\newcount\numfig
\gdef\profonditastruttura{\dp\strutbox}

\def\senondefinito#1{\expandafter\ifx\csname#1\endcsname\relax}

\def\SIA #1,#2,#3 {\senondefinito{#1#2}%
\expandafter\xdef\csname #1#2\endcsname{#3}\else
\write16{???? ma #1,#2 e' gia' stato definito !!!!} \fi}

\def\etichetta(#1){(\veroparagrafo.\veraformula)%
\SIA e,#1,(\veroparagrafo.\veraformula) %
\global\advance\numfor by 1%
\write15{\string\FU (#1){\equ(#1)}}%
\write12{\string\FU (#1){\equ(#1)}}%
\write16{ EQ #1 ==> \equ(#1)  }}

\def\FU(#1)#2{\SIA fu,#1,#2 }

\def\etichettaa(#1){(A\veroparagrafo.\veraformula)%
\SIA e,#1,(A\veroparagrafo.\veraformula) %
\global\advance\numfor by 1%
\write15{\string\FU (#1){\equ(#1)}}%
\write12{\string\FU (#1){\equ(#1)}}%
\write16{ EQ #1 ==> \equ(#1)  }}

\def\getichetta(#1){Fig. \verafigura
\SIA e,#1,{\verafigura} %
\global\advance\numfig by 1%
\write15{\string\FU (#1){\equ(#1)}}%
\write12{\string\FU (#1){\equ(#1)}}%
\write16{ Fig. \equ(#1) ha simbolo  #1  }}

\newdimen\gwidth

\def\BOZZA{
\def\alato(##1){
 {\vtop to \profonditastruttura{\baselineskip
 \profonditastruttura\vss
 \rlap{\kern-\hsize\kern-1.2truecm{$\scriptstyle##1$}}}}}
\def\galato(##1){ \gwidth=\hsize \divide\gwidth by 2
 {\vtop to \profonditastruttura{\baselineskip
 \profonditastruttura\vss
 \rlap{\kern-\gwidth\kern-1.2truecm{$\scriptstyle##1$}}}}}
}

\def\alato(#1){}
\def\galato(#1){}

\def\veroparagrafo{\number\numsec}\def\veraformula{\number\numfor}
\def\verafigura{\number\numfig}

\def\geq(#1){\getichetta(#1)\galato(#1)}
\def\Eq(#1){\eqno{\etichetta(#1)\alato(#1)}}
\def\eq(#1){\etichetta(#1)\alato(#1)}
\def\Eqa(#1){\eqno{\etichettaa(#1)\alato(#1)}}
\def\eqa(#1){\etichettaa(#1)\alato(#1)}
\def\eqv(#1){\senondefinito{fu#1}$\clubsuit$#1
\write16{#1 non e' (ancora) definito}%
\else\csname fu#1\endcsname\fi}
\def\equ(#1){\senondefinito{e#1}\eqv(#1)\else\csname e#1\endcsname\fi}

\openin14=\jobname.aux \ifeof14 \relax \else
\input \jobname.aux \closein14 \fi
\openout12=\jobname.aux

\newcount\pinclude \newcount\pcount
\def\include#1{\pinclude=0
\openin13=#1.aux \ifeof13 \relax \else \input #1.aux \closein13 \fi
\openout15=#1.aux
\input #1
\pcount=\count0 \advance\pcount by 1
\write15{\string\ifnum%
\string\pinclude =1 %
\string\pageno = \number\pcount \string\fi}
\closeout15 }

\def\st{\scriptstyle}

\def\media#1{{\langle#1\rangle}} \def\ie{\hbox{\it i.e.\ }}
  \let\ig=\int \let\io=\infty  \let\i=\infty

\let\dpr=\partial \def\V#1{\vec#1}   
   
\def\xx{{\V x}} \def\yy{{\V y}} \def\kk{{\V k}}

\def\tende#1{\vtop{\ialign{##\crcr\rightarrowfill\crcr
              \noalign{\kern-1pt\nointerlineskip}
              \hskip3.pt${\scriptstyle #1}$\hskip3.pt\crcr}}}
\def\otto{{\kern-1.truept\leftarrow\kern-5.truept\to\kern-1.truept}}

\def\R{{\bf R}}
\def\mbox{\hbox}
\def\Tr{{\rm Tr}\,}\def\EE{{\cal E}}

\def\RR{{\cal R}} \def\NN{{\cal N}} 
  \def\LL{{\cal L}} \def\DD{{\cal D}}
\def\fra#1#2{{#1\over#2}}
\def\ap{{\it a priori\ }}

\def\={{\equiv}}
\def\virg{\qquad,\qquad}
\def\*{\vglue0.3truecm}
\let\0=\noindent \let\\=\noindent

 \def\ST{\tilde S}
\def\undx{{\underline x}} \def\undy{{\underline y}}

\def\guida{\leaders\hbox to 1em{\hss.\hss}\hfill}

\def\indica{\leaders \hbox to 0.5cm{\hss.\hss}\hfill}

\def\initfig#1{
\catcode`\%=12\catcode`\}=12\catcode`\{=12
              \catcode`\<=1\catcode`\>=2
\openout13=#1.ps}
\def\endfig{
\closeout13
\catcode`\%=14\catcode`\{=1
\catcode`\}=2\catcode`\<=12\catcode`\>=12}

\initfig{ini}
\write13</fx [>
\write13<     0.00    1.11    2.22    3.33    4.44    5.56    6.67    7.78>
\write13<     8.89   10.00   11.11   12.22   13.33   14.44   15.56   16.67>
\write13<    17.78   18.89   20.00   21.11   22.22   23.33   24.44   25.56>
\write13<    26.67   27.78   28.89   30.00   31.11   32.22   33.33   34.44>
\write13<    35.56   36.67   37.78   38.89   40.00   41.11   42.22   43.33>
\write13<    44.44   45.56   46.67   47.78   48.89   50.00   51.11   52.22>
\write13<    53.33   54.44   55.56   56.67   57.78   58.89   60.00   61.11>
\write13<    62.22   63.33   64.44   65.56   66.67   67.78   68.89   70.00>
\write13<    71.11   72.22   73.33   74.44   75.56   76.67   77.78   78.89>
\write13<    80.00   81.11   82.22   83.33   84.44   85.56   86.67   87.78>
\write13<    88.89   90.00   91.11   92.22   93.33   94.44   95.56   96.67>
\write13<    97.78   98.89  100.00>
\write13<] def>
\write13</fy [>
\write13<     0.00    2.07    3.98    5.58    6.75    7.39    7.46    6.96>
\write13<     5.91    4.41    2.57    0.52   -1.56   -3.52   -5.21   -6.50>
\write13<    -7.28   -7.50   -7.14   -6.22   -4.82   -3.05   -1.04    1.04>
\write13<     3.05    4.82    6.22    7.14    7.50    7.28    6.50    5.21>
\write13<     3.52    1.56   -0.52   -2.57   -4.41   -5.91   -6.96   -7.46>
\write13<    -7.39   -6.75   -5.58   -3.98   -2.07    0.00    2.07    3.98>
\write13<     5.58    6.75    7.39    7.46    6.96    5.91    4.41    2.57>
\write13<     0.52   -1.56   -3.52   -5.21   -6.50   -7.28   -7.50   -7.14>
\write13<    -6.22   -4.82   -3.05   -1.04    1.04    3.05    4.82    6.22>
\write13<     7.14    7.50    7.28    6.50    5.21    3.52    1.56   -0.52>
\write13<    -2.57   -4.41   -5.91   -6.96   -7.46   -7.39   -6.75   -5.58>
\write13<    -3.98   -2.07    0.00>
\write13<] def>
\write13<>
\write13<>
\write13</cambio_coordinate{ 
\write13<4 copy exch pop exch sub  
\write13<6 1 roll exch pop sub    
\write13<4 1 roll translate       
\write13<2 copy exch atan rotate  
\write13<2 exp exch 2 exp add sqrt  
\write13<} def>
\write13<>
\write13<>
\write13</nx fx length 1 sub def>
\write13<>
\write13</normonda { 
\write13<100 div dup scale>
\write13<fx 0 get fy 0 get moveto>
\write13<1 3 nx {>
\write13<dup dup 1 add exch 2 add 3 1 roll exch>
\write13<dup fx exch get 4 1 roll fy exch get 3 1 roll>
\write13<dup fx exch get 3 1 roll fy exch get exch>
\write13<dup fx exch get exch fy exch get curveto>
\write13<} for } def>
\write13<>
\write13</onda { gsave 
\write13<cambio_coordinate 
\write13<normonda stroke grestore } def>
\write13<>
\write13</freccia { gsave 
\write13<cambio_coordinate 
\write13<dup 0 0 moveto 0 lineto 
\write13<2 div 0 translate>
\write13<15 rotate 0 0 moveto -5 0 lineto -30 rotate 0 0 moveto -5 0 lineto>
\write13<stroke grestore } def>
\write13<>
\write13</punto { gsave  
\write13<2 0 360 newpath arc fill stroke grestore} def>
\write13<>
\write13</cerchio { gsave 
\write13<0 360 newpath arc stroke grestore} def>
\write13<>
\write13</tlinea { gsave 
\write13<moveto [4 4] 2 setdash lineto stroke grestore} def>
\write13<>
\write13</normarco { gsave newpath 
\write13<dup dup 2 div exch  3 sqrt -2 div mul 
\write13<3 copy 3 -1 roll add 
\write13<5 2 roll 3 -1 roll 60 120 arc 
\write13<translate 15 rotate 0 0 moveto -5 0 lineto -30 rotate 0 0>
\write13<moveto -5 0 lineto>
\write13<stroke grestore>
\write13<} def>
\write13<>
\write13</cfreccia { gsave 
\write13<cambio_coordinate 
\write13<normarco grestore } def>
\write13<>
\write13</cfrecciaspe { gsave 
\write13<cambio_coordinate 1 -1 scale 
\write13<normarco grestore } def>
\write13<>
\write13</normarcofin { gsave newpath 
\write13<dup dup dup 2 div exch  3 sqrt -2 div mul 
\write13<3 -1 roll 60 120 arc 
\write13<0 translate -15 rotate 0 0 moveto -5 0 lineto -30 rotate 0 0>
\write13<moveto -5 0>
\write13<lineto stroke grestore>
\write13<} def>
\write13<>
\write13<>
\write13</cfrecciafin { gsave 
\write13<cambio_coordinate 
\write13<normarcofin grestore } def>
\write13<>
\write13</normzigzag { 
\write13<100 div dup scale 0 0 moveto>
\write13<4 { 6.25 7.5 rlineto 12.5 -15 rlineto 6.25 7.5 rlineto} repeat } def>
\write13<>
\write13</zigzag { gsave 
\write13<cambio_coordinate 
\write13<normzigzag stroke grestore } def>
\write13<>
\write13</slinea { gsave 
\write13<setlinewidth 4 2 roll moveto lineto stroke grestore} def>
\write13<>
\write13</sfreccia { gsave 
\write13<setlinewidth freccia grestore} def>
\write13<>
\write13</semicerchio { gsave 
\write13<0 180 newpath arc stroke grestore} def>
\write13<>
\write13</puntino { gsave  
\write13<0.1 0 360 newpath arc fill stroke grestore} def>
\write13<>
\write13</mazza {gsave  
\write13<currentpoint translate rotate 0 0 moveto dup 0 lineto>
\write13<stroke newpath 5 add 0 5 0 360 arc stroke grestore} def>
\write13<>
\write13</gmazza {gsave  
\write13<currentpoint translate rotate 0 0 moveto dup 0 lineto>
\write13<stroke newpath 10 add 0 10 0 360 arc stroke grestore} def>
\write13<>
\write13</ovale {gsave  
\write13<translate rotate 1 .75 scale 0 0 3 2 roll 0 360 newpath arc stroke>
\write13<grestore} def>
\write13<>
\write13</arco {gsave 
\write13<cambio_coordinate 
\write13<newpath dup dup 2 div exch  3 sqrt -2 div mul 
\write13<3 -1 roll 60 120 arc stroke grestore} def>
\endfig

\font\large=cmbx12
\vglue1.truecm
\centerline{\large \noindent Renormalization group approach to zero
temperature Bose condensation.}
\vglue1.truecm
\centerline{\bf G. Benfatto}
\vglue.3truecm
\centerline{Dipartimento di Matematica}
\centerline{Universit\`a di Roma ``Tor Vergata''}
\centerline{Via della Ricerca Scientifica}
\centerline{00133 Roma, Italy}
\vglue.5truecm
\centerline{\bf Talk given at the workshop}
\centerline{\bf ``Constructive results in Field Theory, Statistical Mechanics
and Condensed Matter Physics''}
\centerline{\bf Palaiseau, July 25-27, 1994}
\vglue2.truecm

{\it \S1 Introduction.}
\vskip0.5truecm
\numsec=1\numfor=1\pgn=1

In this lecture, I will present some recent results [B] about the
renormalization group approach to the problem of Bose condensation at zero
temperature for a three dimensional system of bosons, interacting with a
repulsive short range potential.

Before starting, I have to acknowledge the contribution of many other people
in the different stages of this work.  In fact, I was introduced to this
problem by G. Gallavotti and J. F. Perez two years ago, but at the beginning
we were only able to understand the difficulties, without finding a solution,
so that we gave up the project.  I started again to work on the subject during
a visit to the University of Sao Paulo, where I had many useful discussions
with J. F. Perez; these discussions allowed to clarify the problem, that I
could really understand only recently.  In this last stage the comments and
the suggestions of G. Gallavotti had again an important role, as well as the
criticism of C. Castellani, C. Di Castro and M. Grilli, which allowed me to
find and correct an important mistake in my calculations.

The ideas that I shall discuss in this lecture are not yet sufficient for a
complete rigorous treatment of the Bose condensation problem, but they are
formulated in a way, that makes reasonable the conjecture that such a
treatment is possible. The aim of my talk will be to give
convincing arguments that the usual picture of three dimensional Bose
condensation (see, for example,  [ADG], [P]) can be explained in terms of a
renormalization group flow, showing asymptotic freedom and anomalous behaviour
of the two-point correlation function $S(x-y)$, at least order by order in the
running coupling constants.

As we shall see, this anomalous behaviour is related to the fact the Fourier
transform of $S(x)$ has a singularity of the type $(k_0^2+ v^2
\kk^2)^{-1}$, to compare with the singularity $(-ik_0+\kk^2/2m)^{-1}$ of the
free Bose gas. This anomaly explains, according to Landau's criterion [LaL],
the superfluid properties of the system, whose spectrum is expressed, for
small momenta, in terms of collective excitations with speed $v$.

The main missing point, in order to get a rigorous proof, is a suitable
cluster expansion, similar to the one used to analyze the infrared $\phi^4_4$
problem [GK]; this would solve the {\it large field problem}. However, I do
not expect this is a trivial generalization of known techniques, since in the
Bose gas problem one has to use complex gaussian measures, instead of positive
gaussian measures, and this would introduce new technical problems.

There is of course a huge physical literature on the subject of Bose
condensation at zero or small temperature.  As far as I know, the more
convincing results about the superfluid behaviour at zero temperature are
contained in [NN] and [PS] (see also references therein), where the authors
use arguments similar to the ones that I will explain (C.  Castellani made me
aware of this point).  However, at my knowledge, there is no treatment of the
problem explicitly based on renormalization group arguments.

In the next section I will give the relevant definitions and I will discuss
the so called {\it Bogoliubov approximation} [Bo], which is the starting point
of the usual picture of Bose condensation.  In \S3 I will introduce the
renormalization group transformation and the associated effective potentials
and running coupling constants.  Finally, in \S4 I will discuss the flow
equations for the running coupling constants, truncated to second order,
showing that they can solved; the solution implies that the theory is
asymptotically free with a superfluid two-point correlation.

\vskip2truecm

\vskip2.truecm
{\it \S2 The Bogoliubov approximation.}
\vskip0.5truecm
\numsec=2\numfor=1\pgn=1

The problem is the following. Let:
$$ H=\sum\limits_{i=1}^N\left(-{\Delta_{\xx_i}\over2m}-\m\right) +
\l\sum\limits_{i<j} v(\xx_i-\xx_j) \Eq(2.1)$$
be the hamiltonian describing a system of $N$ bosons in $\R^3$,
enclosed in a periodic box of side size $L$, interacting with a pair
potential $\l v(\xx-\yy)$ which is supposed $C^\infty$ and with short range
$p_0^{-1}$. In fact
we shall suppose that the interaction is repulsive in the sense that $v(0)>0$,
$v(\xx)\ge 0$ and $\l\ge 0$.

Let $\f^\pm_\xx$ be the creation and the annihilation operators for the
bosons and
$$\f^\s_x=e^{H t}\f^\s_\xx e^{-H t} \quad,\quad \s=\pm,\,
x=(t,\xx)\Eq(2.2)$$
Define, for $\b> t_i \ge 0$, $i=1,\ldots,n$, with $t_i\not=
t_j$ for $i\not=j$:
$$ S_{\s_1\cdots\s_n} (x_1,\ldots,x_n) = \lim_{\b\to\io} \lim_{L\to\io}
{\Tr \left[ e^{-\b H} \f^{\s_{\p(1)}}_{x_{\p(1)}} \cdots
\f^{\s_{\p(n)}}_{x_{\p(n)}}\right] \over \Tr\; e^{-\b H}} \Eq(2.3)$$
where $\s_i=\pm1$ and $\p$ is the permutation of $(1,\ldots,n)$, such that
$t_{\p(1)} > t_{\p(2)} > \ldots > t_{\p(n)}$.
In particular consider:
$$ S(x)=S_{-+}(x,0) \Eq(2.4)$$
The functions \equ(2.3), \equ(2.4) describe the properties of the
{\it ground state} of the above bosons system (essentially {\it by
definition}) in the grand canonical ensemble with chemical potential
$\m$.

The case $\l=0$ is trivial and one finds that, if $\m<0$, the Schwinger
functions at finite $\b$ and $L$ (the r.h.s. of \equ(2.3)) can be calculated
by the Wick rule (see for example [NO]) with propagator:

$$S_{\b,L}(x)=L^{-d}\sum_\kk e^{-i\kk\cdot\xx }e^{-\e(\kk)t}
\Big({\th(t>0) \over 1-e^{-\b\e(\kk)} } +
{\th(t\le 0) e^{-\b\e(\kk)} \over 1-e^{-\b\e(\kk)}}\Big)\Eq(2.5)$$
where
$$\e(\kk)={\kk^2\over 2m}-\m\Eq(2.6)$$

Note that it is not possible to take $\m=0$ in \equ(2.5), since in this case
the term in the sum with $k_0=|\kk|=0$ involves a division by zero.
However the interesting phenomena just
occur when $\m=0$ and we shall therefore have to deal also with such case.
For this reason we take $\m$ as a function of $\b,L$, which goes to $0$ as
$L,\b\to\io$, in such a way that the number of particles in the {\it
condensate} (the state $\kk=0$) is fixed, that is:
$$e^{\b\m}(1-e^{\b\m})^{-1} = L^{d}\r \Eq(2.7)$$
where $\r$ is the condensate density. One finds, if $x_0=t$:
$$S(x)= S_\r(x) \equiv \r+\fra1{(2\p)^3}\ig d\kk
e^{-i\kk\cdot\xx}e^{- {\kk^2\over 2m} x_0}\th(x_0>0) \Eq(2.8)$$
so that:
$$S_0(x) = \fra1{(2\p)^4}\ig dk\,\fra{e^{-ikx}}{-i k_0+ {\kk^2\over 2m}}
\Eq(2.9)$$
where $k=(k_0,\kk)$.

The Schwinger functions generated by the above limiting procedure and by
the Wick rule will describe the ground state of a system
of non interacting bosons with density $\r$. They describe a
{\it Bose condensed state}.

Let us now consider the case $\l>0$. As it is well known [NO], if $\b$ and $L$
are finite, the Schwinger functions can be expressed as functional integrals
in the following way:

$$S_{\s_1\ldots\s_n}(x_1\ldots x_n)=\ig \f^{\s_1}_{x_1}\ldots\f_{x_n}^{\s_n}
{e^{-V(\f)} P(d\f)\over \ig e^{-V(\f)} P(d\f)}\Eq(2.10)$$
where $P(d\f)$ is a complex gaussian measure, such that the
fields $\f^-_x$, $\f^+_x=(\f^-_x)^*$ have covariance:
$$\eqalign{
\ig \f^-_x \f^+_y P(d\f) &= S_{\b,L}(x-y)\cr
\ig \f^-_x \f^-_y P(d\f) &= \ig \f^+_x \f^+_y P(d\f) =0\cr} \Eq(2.11)$$
and, if $x=(x_0,\xx)$, $y=(y_0,\yy)$,
$\L=[-\fra12\b,\fra12\b]\times[-\fra12L,\fra12L]^3$:
$$V(\f)= \l\ig_\L
v(\xx-\yy)\d(x_0-y_0)\f^+_x\f^-_x\f^+_y\f^-_y \,dx\,dy \Eq(2.12)$$
Note that the fields $\f_x^\pm$ satisfy periodic boundary conditions in $\L$.

In order to study the limit $L,\b\to\i$, it is convenient to add to $V(\f)$ a
term proportional to $\ig\f^+_x\f^-_x\, dx$, by changing the definition of
$\m$, and then
fix $\m$ so that \equ(2.7) is satisfied. The limiting theory will be still
described by the functional integrals \equ(2.10), where $P(d\f)$ is the
formal integration with propagator \equ(2.8) and
$$V(\f)= \l\ig_\L v(\xx-\yy)\d(x_0-y_0)\f^+_x\f^-_x\f^+_y\f^-_y \,dx\,dy
+\n \ig_\L \f^+_x\f^-_x\, dx \Eq(2.13)$$

$\n$ has the role of a control parameter that should be fixed so that the
limiting theory is meaningful as a perturbation of the free theory with
propagator \equ(2.8).  If this program is successful, it is natural to say
that there is Bose condensation at $T=0$ with given density $\r$ and chemical
potential $-\n$.  One could also fix the physical mass of the particles by
adding to $V(\f)$ a term $\a \ig \f^+_x (-\D_\xx/ 2m) \f^-_x\, dx$, but we
shall not do that, because we do not expect that the choice of $\a$ is
critical.

The form of the covariance \equ(2.8) shows that the field $\f^\pm_x$
can be represented as:
$$\f_x^\pm=\x^\pm+\psi^\pm_x\Eq(2.16)$$
with $\x^\pm$ being variables independent from $\psi^\pm_x$ and with
covariance $\media{\x^-\x^+}=\r$, while the fields $\psi^\pm_x$ have
covariance $\media{\psi^-_x \psi^+_y} = S_0(x-y)$.

The integration with respect to $\x^\pm$ can be thought as a gaussian
integral by writing $\x^\pm=\x_1\pm i \x_2$ and:
$$P(d\x)={e^{-\fra{\x_1^2 + \x_2^2}{\r} } }
\fra{d\x_1\,d\x_2}{2\p\r}\Eq(2.17)$$
Hence, if we define:
$$W^{(-\io)}(\x) = -{1\over \L} \log \ig e^{-V(\xi + \psi)} P(d\psi)
\Eq(2.18)$$
we see that the computation of
$\media{\x^+\x^-}$ in presence of interaction will lead to the integral:
$$\r=\ig \fra{d\x_1 d\x_2}{2\p\r} (\x_1^2+\x_2^2) \,e^{-(\x_1^2+\x_2^2)/\r}\,
e^{-(W^{(-\io)}(\x)-W^{(-\io)}(0))|\L|}\Eq(2.19)$$
and the equality to $\r$ of the above integral is just the requirement
that the condensate density should be $\r$. Therefore equality
\equ(2.19) can hold if and only if the function $W^\io(\x)$, which is a
function of the product $\x^+\x^-$, by symmetry considerations, reaches
its minimum at $\x^+\x^-=\r$. And in this case $\x^+\x^-$ will be a sure
random variable, provided the minimum is not degenerate.

This implies that, in order to get a condensate with density $\r$, one has to
choose $\n$ so that the free energy \equ(2.18) has a minimum in

$$\x^+=\x^-=\sqrt{\r} >0 \Eq(2.20)$$

These considerations are a heuristic justification of the so called {\it
Bogoliubov approximation}, very usually used in the literature, consisting in
replacing the fields $\x^\pm$ by a real positive constant external field and
by choosing its value so that the ground state energy is minimum.

At my knowledge, there is only one rigorous general result in this direction,
found by Ginibre in 1967, [Gi].  Ginibre proved that, at any finite
temperature and any fixed chemical potential, the finite volume pressure of
the Bose system in the Bogoliubov
approximation has a supremum (as a function of $\x$), whose thermodynamical
limit coincides with the thermodynamical limit of the real pressure. However,
Ginibre was not able to prove a similar result for the correlation functions.

In any case, I will suppose that the Bogoliubov approximation is correct in
the limit $\b,L\to \io$ and that the equation relating $\n$ and $\r$ is
invertible; this implies that,
in order to study Bose condensation at $T=0$, one has to consider the measure
$${1\over \NN} e^{-V_\r(\psi)} P(d\psi) \Eq(2.21a)$$
where $\NN$ is a normalization constant, $V_\r(\psi)$ is obtained from
\equ(2.13) trough the substitution
\equ(2.16), $\x^\pm$ are positive constants satisfying \equ(2.20) and $\n$
has to be chosen so that the free energy is minimum for the fixed value of
$\x^\pm$.

The Schwinger functions of the field $\psi$ are defined by an expression
similar to \equ(2.10) and we shall use the symbol $\ST$ to denote them.
Their perturbation expansion is obtained in the usual way in terms of the
propagator $S_0(x-y)$.
Note that the measure \equ(2.21a) does not preserve the free measure property
$\ST_{--}(x-y)=\ST_{++}(x-y)=0$.

An old perturbative argument [HP] allows to show (see also [ADG]) that the
free energy is minimum if the following formal equation is satisfied:

$$\Si_{-+}(0)=\Si_{++}(0) \Eq(2.22)$$
where $\Si_{\s_1\s_2}(k)$ is the Fourier transform of the sum of all one
particle irreducible graphs (connected graphs which can not become
disconnected by cutting one leg) with two external lines $\psi^{\s_1}_x$,
$\psi^{\s_2}_y$.

The formal proof of \equ(2.22) is very simple. It starts from the observation
that:
$$W^{(-\io)}(\x) = \sum_{s=0}^\io w_s \r^s \Eq(2.23)$$
where $w_s$ is the sum of all irreducible graphs with $2s$ zero momentum
$\x$ external legs ($s$ of type $+$ and $s$ of type $-$), constructed by the
interaction \equ(2.13) and the propagator $S_{\b,L}-\r$ (whose limit for
$\L\to\io$ is given by \equ(2.9)); the irreducibility condition follows from
the fact that the propagator is zero at $k=0$. Hence the free energy is
stationary (and it turns out that the stationary point is unique and a
minimum, at least for $\l$ small enough) if the following equation is
satisfied:
$$\sum_{s=1}^\io s w_s \r^{s-1} =0 \Eq(2.24)$$

Let us now consider $\Si_{-+}(0)$. It is clear that, given a graph
contributing to $w_s$, there are $s^2$ graphs contributing to $\Si_{-+}(0)$
and having the same value; they are obtained by substituting in all possible
ways an external leg $\x^-$ with $\psi^-$ and an external leg $\x^+$ with
$\psi^+$, so that:
$$\Si_{-+}(0) = \sum_{s=1}^\io s^2 w_s \r^{s-1} \Eq(2.25)$$

A similar argument can be used for $\Si_{++}(0)$ or $\Si_{--}(0)$, which are
equal by symmetry reasons. Given a graph contributing to $w_s$, there are
$s(s-1)/2$ equal contributions to $\Si_{++}(0)$, obtained by substituting two
external legs $\x^+$ with two $\psi^+$. Moreover there is a factor $2$, coming
from the possibility of interchanging the two external $\psi^+$ fields in the
calculation of $\Si_{++}(0)$; hence:
$$\Si_{++}(0) = \sum_{s=1}^\io s(s-1) w_s \r^{s-1} \Eq(2.26)$$

The condition \equ(2.22) immediately follows from \equ(2.24), \equ(2.25) and
\equ(2.26). Note also that we could repeat the previous arguments in presence
of an ultraviolet cut off, that we shall indeed introduce in the following
section.

The {\it renormalization condition} \equ(2.22) has here the same role of the
condition which determines the critical temperature in Statistical Mechanics;
for example, in the case of the $\f^4_4$ infrared model, a similar condition
allows to define the stable manifold of the trivial fixed point in the
renormalization group flow [GK].  Therefore it is natural to use it, instead
of the free energy minimum condition, as the condition fixing the chemical
potential, in a renormalization group analysis of the measure \equ(2.21a).

\vskip2truecm

\vskip2.truecm
{\it \S3 The effective potentials.}
\vskip0.5truecm
\numsec=3\numfor=1\pgn=1

The problem we want to study is an infrared problem; therefore we shall
consider a simplified model by substituting $S_0(x)$ with
$$g_{\le 0}(x)= \fra1{(2\p)^4}\ig dk\,t_0(k) \fra{e^{-ikx}}{-i
k_0+ {\kk^2\over 2m}} \Eq(2.14)$$
where $t_0(k)$ is a smooth function, which imposes an ultraviolet cutoff on
scale $p_0$ (the scale of the potential). We shall choose $t_0(k)$ as a
regularization of the characteristic function of the set
$\{k_0^2 +{\kk^2\over 2m} {p_0^2\over 2m}
\le ({p_0^2\over 2m})^2\}$, such that, if $\g$ is a fixed
positive number greater than $1$ and $q_0=p_0^2/2m$:
$$t_0(k) = \theta(k_0^2 +{\kk^2\over 2m} {p_0^2\over 2m}) \quad,\quad
\theta(\l) = \cases{ 1 & if $|\l|
\le {2\g^2\over \g^2+1} q_0^2$\cr 0 & if $|\l| \ge {2\over \g^2+1} q_0^2$\cr}
\Eq(2.15)$$
The particular choice of $\theta(\l)$ is not relevant; it is done only to
simplify some calculations in the following.

Note that the assumed presence of the ultraviolet cut off on the scale of the
interaction potential is reasonable only if $\r p_0^{-3}\ll 1$, \ie only
if there is in mean much less than one particle in a cube with side equal to
the range $p_0^{-1}$ of the potential.

Hence we have to study the measure:
$${1\over \NN} e^{-V_\r(\psi)} P^{(\le 0)}(d\psi) \Eq(2.21)$$
where $P^{(\le 0)}(d\psi)$ is the measure with covariance \equ(2.14).
We shall do that by a multiscale analysis, in the form
presented in [G] and applied to a ``similar'' infrared problem, the one
dimensional Fermi gas, in [BG], [BGPS].

At first, one could try to define a family of effective potentials based on a
multiscale decomposition of the covariance \equ(2.14).  However, this
approach does not work, as a consequence of the well known fact that the
functions $\ST_{\s_1\s_2}(x-y)$ are expected to have a large distance
behaviour very different from the behaviour of the {\it free propagator}
\equ(2.14).  In fact, by simple (very formal) perturbative arguments,
using the normalization condition \equ(2.22), one can
prove [ADG] that the Fourier transform of $\ST_{\s_1\s_2}(x-y)$ behaves for
$k\to 0$ as

$${A\over k_0^2 + c^2 \kk^2} \Eq(3.1)$$
where $A$ and $c$ are suitable constants ($c$ has the physical
interpretation of sound waves velocity in the condensate). The behaviour
\equ(3.1) is considered typical of superfluids.

Hence we are faced with a model with an {\it anomalous} behaviour, as in the
$d=1$ Fermi gas. The anomaly is now of a different nature, but we shall see
that we can extend in a very natural way the renormalization group techniques
of [BG] to cover also this case.

We start writing the potential $V_\r$ in the following way:
$$V_\r = |\L| (\n\r + \l\hat v(0)\r^2) + \LL V_\r + \RR V_\r
\Eq(3.2)$$
where the {\it local part} $\LL V_\r$ is obtained by substituting
$\psi^\pm_y$ with $\psi^\pm_x$ in the expression of $V_\r(\psi)$, after
extracting the constant term, so that:
$$\eqalign{
&\LL V_\r(\psi)=\l\hat v(0)\ig_\L (\psi^+_x\psi^-_x)^2\,dx+
2\l\hat v(0)\sqrt\r\ig_\L\psi^+_x\psi^-_x\,(\psi^+_x+\psi^-_x)\,dx+\cr
&+(2\l\hat v(0)\r+\n)\ig_\L\psi^+_x\psi^-_x\,dx+
\l\hat v(0)\r \ig_\L (\psi^+_x + \psi^-_x)^2 \,dx \cr}\Eq(3.3)$$

We now observe that, if the ``remainder'' $\RR V_\r$ and the local terms of
order greater than two were not present in \equ(3.3) (so obtaining the
exactly soluble {\it Bogoliubov model}), then the condition
\equ(2.22) could be imposed, by choosing $\n=-2\l\r\hat v(0)$, that is by
putting equal to zero the coefficient of $\psi^+\psi^-$ in \equ(3.2).
This observation implies that the ``natural'' way to study the measure
\equ(2.21) is to change the free measure, by absorbing in it the local terms
of $V_\r$, quadratic in the field, which remain after imposing the
condition $\n^0 \equiv \n+2\l\r\hat v(0)=0$. We shall denote $P_B(d\psi)$ the
new measure and we shall call it the {\it renormalized free measure}.

By using the results of Appendix 1 with $a=\l\r\hat v(0)$, $b_0=b=c=0$, we see
that the renormalized free measure has indeed the behaviour \equ(3.1) for
$k\to 0$, with $c^2=c_B^2 \equiv 2\l\r\hat v(0)$.

Therefore the measure \equ(2.21) should be written:
$${1\over \NN} e^{-\tilde V_0(\psi)}P_B(d\psi)\Eq(3.4)$$
where
$$P_B(d\psi)\=P^{(\le 0)}(d\psi)e^{-\l\hat v(0)\r\ig_\L
(\psi^+_x+\psi^-_x)^2\ dx}\Eq(3.5)$$
and
$$\eqalign{
\tilde V_0(\psi)=&\l\hat v(0)\ig_\L(\psi^+_x\psi^-_x)^2\,dx+\cr
&+2\l\hat v(0)\sqrt\r\ig_\L\psi^+_x\psi^-_x(\psi^+_x+\psi^-_x)\,dx
+\n^0\ig_\L \psi^+_x\psi^-_x\,dx\cr}\Eq(3.6)$$
and the parameter $\n^0$ has to be determined so that the
condition \equ(2.22) is satisfied.

It is convenient, before proceeding, to change the basic fields and to
perform a rescaling (amounting at fixing $\r=1$); we set:
$$\chi^\pm_x = \fra1{\sqrt{2\r}}(\psi_x^+\pm\psi_x^-),\qquad
\psi^{\pm}=\sqrt{\fra\r2}\,(\chi_x^+\pm\chi_x^-) \Eq(3.7)$$
$$\e = \l\hat v(0)\r\fra{2m}{p_0^2} \Eq(3.7a)$$
so that $\tilde V_0$ becomes a function of $\chi$:
$$\eqalign{
&\tilde V_0(\chi)=\fra{p_0^2\r}{2m}\Big(\fra\e4\ig_\L
((\chi^-_x)^4-2(\chi^-_x)^2(\chi^+_x)^2+(\chi^+_x)^4)dx+\cr
&+\e\sqrt2\ig_\L((\chi^+_x)^3-(\chi^-_x)^2\chi^+_x)\,dx+
\fra{\n^0}2\fra{2m}{p_0^2}\ig_\L
((\chi^+_x)^2 -(\chi^-_x)^2)\,dx\Big)\cr}\Eq(3.8)$$
and the covariance of the fields $\chi^\pm$ in the distribution
\equ(3.5) is, in the limit $\L\to\io$:
$$\tilde C^{\s_1\s_2}(x-y) = \int P_B(d\c) \c_x^{\s_1} \c_y^{\s_2} =
\fra1{(2\p)^4}\ig e^{-i k(x-y)} t_0(k) \tilde G_0^{-1}(k)_{\s_1\s_2}
\Eq(3.9)$$
where the matrix $\tilde G_0(k)$ is defined by (see appendix 1):
$$\tilde G_0(k)=\r \pmatrix{
\fra{\kk^2}{2m}+4\e\fra{p_0^2}{2m} t_0(k)&ik_0\cr
-ik_0&-\fra{\kk^2}{2m}\cr}\Eq(3.10)$$
where the first row and column correspond to $\s=+$ and the second row
and column correspond to $\s=-$.

Note that $\tilde C^{--}(k)$ behaves, for $k\to 0$, as $1/k^2$ (like in
$\f^4_4$
theory), while $\tilde C^{-+}(k)$ and $\tilde C^{++}(k)$ are less singular. In
a sense, the field $\c^+$ behaves like the derivative of the field $\c^-$.

In order to study the measure \equ(3.4), we make the scale decomposition
$$\tilde C^{\s_1\s_2}(x) = \sum_{h=-\i}^0 \tilde C_h^{\s_1\s_2}(x) \Eq(3.11)$$
where $\tilde C_h^{\s_1\s_2}(x)$ is obtained from \equ(3.9) by substituting
$t_0(k)$ with
$$\tilde T_h(k) = t_0(\g^{-h}k) -t_0(\g^{-h+1}k) \Eq(3.12)$$
so that we have the scaling relations:
$$\tilde C_h^{\s_1\s_2}(x) = \g^{(3+{\s_1+\s_2\over 2})h}
\tilde g_h^{\s_1\s_2}(\g^h x) \Eq(3.13)$$
with
$$\tilde g_h^{\s_1\s_2}(x) =
{1\over (2\p)^4} \int dk e^{-ikx} \tilde T_0(k) \tilde G_h^{-1}(k)_{\s_1\s_2}
\Eq(3.14)$$
and
$$\tilde G_h(k)=\r \pmatrix{\fra{\g^{2h} \kk^2}{2m}+4\e\fra{p_0^2}{2m}
t_0(\g^h k) &ik_0\cr -ik_0&-\fra{\kk^2}{2m}\cr}\Eq(3.15)$$
Note that $\tilde g_h^{\s_1\s_2}(x)$ are essentially independent of $h$, for
$h\to -\i$.

Let us now define the {\it effective potentials} $\tilde V_h(\chi)$ in
the usual way [G]:
$$\eqalign{
&e^{-\tilde V_h(\chi)} = \ig \tilde P_B^{(h+1)}(d\chi^{(h+1)}) \ldots
\tilde P_B^{(0)}(d\chi^{(0)})
e^{-\tilde V_0(\chi + \chi^{(h+1)} + \ldots + \chi^{(0)})}=\cr
&=\ig \tilde P_B^{(h+1)}(d\chi^{(h+1)}) e^{-\tilde
V_{h+1}(\chi+\chi^{(h+1)})}\cr}\Eq(3.16)$$
where $\tilde P_B^{(h)}(d\chi)$ denotes the measure with covariance $\tilde
C_h$.

The scaling relations \equ(3.13) imply that we can define dimensionless
fields $\bar\c^\pm$ by the relations:
$$\c^-_x = \g^h\bar\c^-_{\g^hx} \qquad,\qquad \c^+_x = \g^{2h}\bar\c^+_{\g^hx}
\Eq(3.17)$$
and this allows us to analyze the relevant local part $\LL\tilde V_h(\c)$ of
the effective potentials $\tilde V_h(\c)$.

$\LL\tilde V_h(\c)$ is in principle a linear combination of all the local
terms relevant or marginal. However only a few of them are different from
zero, thanks to the symmetries of the original potential \equ(2.13).
Let us consider first the local terms
not involving derivatives of the fields, that is terms of the form
$$F_{m_1 m_2} = \int dx\, {\c^-_x}^{m_1} {\c^+_x}^{m_2} \Eq(3.18)$$
By using \equ(3.17), it is easy to show that $F_{m_1 m_2}$ is not irrelevant
only if $m_1+m_2-4 \le 0$. The terms with $m_1+m_2=1$ are absent, since the
fields $\c^\pm$ have no $k=0$ Fourier component. If $m_1+m_2=2$, we have two
relevant terms, $F_{20}$ and $F_{11}$, and one marginal term, $F_{02}$. If
$m_1+m_2\ge 3$, we have one relevant term, $F_{30}$, and two marginal
terms, $F_{40}$ and $F_{31}$. However, $F_{11}$ and $F_{30}$ are absent,
as a consequence of the particular structure of the
potential \equ(2.13), which implies that the local monomials of order $2$ and
$3$ in the fields $\psi^\pm$ {\it must} appear in the following
combinations:
$$\eqalign{
&(\psi^+_x)^3+(\psi^-_x)^3=2\fra{\r^{3/2}}{2^{3/2}}
((\chi^+_x)^3+3\chi^+\,(\chi^-)^2)\cr
&\psi^+_x\psi^-_x\,(\psi^+_x+\psi^-_x)=2\fra{\r^{3/2}}{2^{3/2}}
((\chi^+_x)^3-(\chi^+_x)(\chi^+_x)^2)\cr
&\psi_x^+\psi_x^-=\fra\r2((\chi^+_x)^2-(\chi^-_x)^2)\cr
&(\psi_x^+)^2+(\psi^-_x)^2=2\fra\r2((\chi^+_x)^2+(\chi^-_x)^2)\cr}\Eq(3.19)$$

As regards the local terms involving derivatives of the field, we have three
marginal terms, that is:
$$\eqalign{
D_{tt} &=-(\fra{2m}{p_0^2})^2\ig_\L (\dpr_{x_0}\chi^-_x)^2\,dx
\quad,\quad
D_{ss}=-p_0^{-2}\ig_\L (\dpr_{\xx}\chi^-_x)^2\,dx \cr
D_t &=- \fra{2m}{p_0^2} \ig_\L \chi^+_x \dpr_{x_0}\chi^-_x\,dx \cr}
\Eq(3.20)$$
We did not include in the list the local term  $-\int_\L dx \chi^+_x \dpr_\xx
\chi^-_x = (1/\L) \sum_k \chi^+_k i\kk$ $\chi^-_k$, since it is identically
zero; in fact the fields $\chi^\s_k$ are even functions of
$\kk$. Finally, the local terms $\ig_\L dx \chi^-_x\dpr\chi^-_x$ and
$\ig_\L dx (\chi^-_x)^2 \dpr\chi^-_x$ are absent, since they are integrals
of total derivatives and the fields satisfy periodic boundary conditions.

We can now define the relevant part of the effective potentials in terms of a
{\it localization operator}, defined in the usual way [G]. For example, if
$(m,n)$ is equal to $(4,0)$, $(2,1)$ or $(0,2)$, we define:
$$\LL \chi^-_{x_1} \ldots \chi^-_{x_m} \chi^+_{y_1} \ldots \chi^+_{y_n} =
(\chi^-_{x_1})^m (\chi^+_{x_1})^n \Eq(3.20a)$$
and a similar definition is used for the other marginal terms.
The monomials of positive dimension
are localized trough a suitable Taylor expansion:
$$\eqalign{
\LL \chi^-_x \chi^-_y &= (\chi^-_x)^2 + \sum_{i=0}^3 (y_i-x_i)\chi^-_x\dpr_i
\chi^-_x +
{1\over 2} \sum_{i,j=0}^3 (y_i-x_i)(y_j-x_j) \chi^-_x \dpr_i\dpr_j \chi^-_x\cr
\LL \chi^+_x \chi^-_y &= \chi^+_x \chi^-_x +
\sum_{i=0}^3 (y_i-x_i) \chi^+_x \dpr_i \chi^-_x\cr} \Eq(3.20b)$$
Note that, according to the remark preceeding \equ(3.19) and rotation
invariance,
many terms in the r.h.s. of \equ(3.20b) cancel out in the full effective
potential.

The previous discussion implies that $\LL\tilde V_h(\c)$ is of the form
$$\eqalign{
\LL\tilde V^{(h)}(\chi)=&\fra{p_0^2\r}{2m}
\big(\tilde \l_h F_{40}+ \tilde \m_h F_{21} +\g^{2h} \tilde \n_h
(F_{02}-F_{20}) +\cr
&+2 \tilde z_h F_{02}+ 2 \tilde \z_h D_{tt} +2 \tilde \a_h D_{ss} + 2
\tilde d_h D_t\big)\cr}\Eq(3.21)$$
where the factors $2$ have no special meaning and the
dimension fixing factor $\fra{p_0^2\r}{2m}$ is introduced to keep track of the
dimensions of the various quantities (its physical dimension is that of
an action density, in space time).

As usual,
we can hope to have a perturbative control of the model only if we can
show that all the running constants appearing in \equ(3.21) stay small for
$h\to-\i$, for a suitable choice of $\n_0$, such that:
$$|\tilde \n_h| \le \bar\n \virg \hbox{for}\ h\to -\io \Eq(3.22)$$

This condition is equivalent to the renormalization condition \equ(2.22). In
fact it implies that, for $h\to-\io$, the effective potential is of the second
order in the fields $\psi^\pm$ and that they appear only in the combination
$(\psi^+_x +\psi^-_x)^2=2\r F_{02}$, up to terms containing field derivatives,
which do not influence \equ(2.22). A simple calculation allows to prove that
this structure of the effective potential implies the renormalization
condition.

For $h=0$ we have:
$$\tilde \l_0=\fra\e4,\quad
\tilde \m_0=-\e\sqrt2,\quad \tilde \n_0=\fra{\n^0}2\fra{2m}{p_0^2} \Eq(3.23)$$
and
$$\tilde z_0=\tilde \z_0=\tilde \a_0=\tilde d_0=0 \Eq(3.24)$$
However $\tilde z_h$, $\tilde \z_h$, $\tilde d_h$ and $\tilde \a_h$ will be
different from zero for $h<0$ and they could grow as $h\to-\i$.

This problem can be solved by the same strategy used in passing from the
representation \equ(2.21) of the measure to the representation \equ(3.4).
We define iteratively a new family of effective potentials $V_h(\c)$ in the
following way.

Given $V_0(\c)=\tilde V_0(\c)$, we define $\tilde V_{-1}(\c)$ as before, so
that:
$$\eqalign{
{1\over \NN} \int P_B(d\c) e^{-V_0(\c)} &=
{1\over \NN} \int \tilde P_B^{(\le-1)}(d\c) \tilde P_B^{(0)}(d\c^{(0)})
e^{-V_0(\c+\c^{(0)})} = \cr
&= {1\over \NN} \int \tilde P_B^{(\le-1)}(d\c) e^{-\tilde V_{-1}(\c)}
\cr } \Eq(3.25)$$
where $\tilde P_B^{(\le-1)}(d\c)$ and $\tilde P_B^{(0)}(d\c)$ denote the
measures
with covariance $\sum_{h=-\i}^{-1} \tilde C_h^{\s_1\s_2}$ and
$\tilde C_0^{\s_1\s_2}$, respectively.

We then define $V_{-1}(\c)$ by absorbing the terms proportional to
$\tilde z_{-1}$,
$\tilde \z_{-1}$, $\tilde d_{-1}$ and $\tilde \a_{-1}$ in the measure
$\tilde P_B^{(\le-1)}(d\c)$, so that
$${1\over \NN} \int \tilde P_B^{(\le-1)}(d\c) e^{-\tilde V_{-1}(\c)} =
{1\over \NN} \int P_B^{(\le-1)}(d\c) e^{- V_{-1}(\c)} \Eq(3.26)$$

We can iterate this procedure, so defining a family of measures
$P_B^{(\le h)}(d\c)$ and a family of effective potentials $V_h(\c)$,
such that:
$$\LL V_h(\c) = \fra{p_0^2\r}{2m} \left[\l_h F_{40} + \m_h F_{21} + \g^{2h}
\n_h (F_{02} - F_{20}) \right] \Eq(3.27)$$
and the covariance $C^{\s_1\s_2}_{\le h}$ of $P_B^{(\le
h)}(d\c)$ is of the form (as follows from the calculations in appendix 1):
$$C^{\s_1\s_2}_{\le h}(x-y) =
{1\over (2\p)^4} \int dk
e^{-ik(x-y)} \, t_0(\g^{-h}k) \,G_{\le h}^{-1}(k)_{\s_1\s_2}
\Eq(3.28)$$
where
$$G_{\le h}(k)=\r
\pmatrix{\fra{\kk^2}{2m}+4\fra{p_0^2}{2m} \bar Z_{h}(k)& ik_0\bar E_{h}(k)\cr
-i k_0\bar E_{h}(k)&-\big(\fra{\kk^2}{2m}+ 4\bar
B_{h}(k) \fra{2m}{p_0^2}\,k_0^2+ 4 \bar A_{h}(k) \fra{\kk^2}{2m})\cr}
\Eq(3.29)$$

Note that $V_h(\c)$ can be easily obtained from the potential $\bar V_h(\c)$
defined by:
$$
e^{-\bar V_{h-1}(\c)} =
{1\over \NN} \int \bar P_B^{(h)}(d\c^{(0)}) e^{-V_h(\c+\c^{(0)})}
\Eq(3.30)$$
where $\bar P_B^{(h)}(d\c)$ is a single scale covariance obtained from
\equ(3.28) by substituting $t_0(\g^{-h}k)$ with
$$T_0(\g^{-h}k) = t_0(\g^{-h}k) -t_0(\g^{-h+1}k) \Eq(3.31)$$
$V_h(\c)$ is calculated by the equation:
$$V_h(\c) = \bar V_h(\c) - \fra{p_0^2\r}{2m} \left[
2 z_h F_{02} +2 \z_h D_{tt} +2 \a_h D_{ss} +2 d_h D_t
\right] \Eq(3.32)$$
if the local part of $\bar V_h$ is written in the form
$$\eqalign{
\LL\bar V_h(\c) = \fra{p_0^2\r}{2m} [ &\l_h F_{40} + \m_h F_{21} +
\g^{2h} \n_h (F_{02} - F_{20}) + \cr
&+ 2z_h F_{02} + 2\z_h D_{tt} + 2\a_h D_{ss}
+2 d_h D_t ] \cr} \Eq(3.33)$$

Note that the choice \equ(2.15) of $t_0(k)$ implies that
$$T_h(k) T_{h'}(k) = 0 \qquad \hbox{if} |h-h'|>1 \Eq(3.34)$$

The functions $\bar A_h(k)$, $\bar B_h(k)$, $\bar Z_h(k)$ and $\bar
E_h(k)$ satisfy, for $h\le -1$, some
recursion relations, which can be easily obtained using \equ(3.32) and the
relation \equ(A1.11) of appendix 1, implying that
$$G_{\le h}(k) = G_{\le h+1} + 2 \D_h(k) t_0(\g^{-h}k)
\Eq(3.35)$$ %
where:
$$\D_h(k) = \r\pmatrix{2  z_h\fra{p_0^2}{2m} & i d_hk_0\cr
-i d_hk_0& -2\fra{2m}{p_0^2} k_0^2
\z_h- 2\a_h\fra{\kk^2}{2m}\cr}\Eq(3.36)$$
It follows that:
$$\eqalign{
\bar Z_h(k) &= \bar Z_{h+1}(k) +  z_h t_0(\g^{-h}k) \cr
\bar A_h(k) &= \bar A_{h+1}(k) + \a_h t_0(\g^{-h}k) \cr
\bar B_h(k) &= \bar B_{h+1}(k) + \z_h t_0(\g^{-h}k) \cr
\bar E_h(k) &= \bar E_{h+1}(k) + 2 d_h t_0(\g^{-h}k) \cr
} \Eq(3.37)$$
with initial conditions:
$$\bar Z_0(k)=\e t_0(k) \qquad,\qquad \bar A_0(k)=\bar B_0(k)=0 \qquad,\qquad
\bar E_0(k)=1 \Eq(3.38)$$

The covariance $\bar C_h^{\s_1\s_2}(x)$ associated with the measure $\bar
P_B^{(h)}(d\c)$ satisfies the following scaling relations:
$$\bar C_h^{\s_1\s_2}(x) = \g^{(3+{\s_1+\s_2\over 2})h}
g_h^{\s_1\s_2}(\g^h x) \Eq(3.39)$$
where
$$g^{\s_1\s_2}_h(x) = {1\over (2\p)^4} \int dk
\fra{e^{-ikx} \; T_0(k) \;  p_h^{\s_1\s_2}(k)}{\r \DD_h(k) } \Eq(3.40)$$
$$\DD_h(k) = b_h(k) k_0^2+ a_h(k) {v_0^2\over 4} \kk^2 +
\g^{2h} [ {\kk^4\over 4m^2} (1 + 4A_h(k) ) + 4 B_h(k) {k_0^2\kk^2\over p_0^2}]
\Eq(3.41)$$
$$\eqalign{a_h(k) &= 4Z_h(k) + 16Z_h(k) A_h(k) \cr
b_h(k) &= E_h(k)^2 +16Z_h(k) B_h(k) \cr}\Eq(3.42)$$
$$\eqalign{
p_h^{-+}(k) &=  p_h^{+-}(-k) = -ik_0E_h(k) \virg
p_h^{--}(k) = -\g^{2h}{\kk^2\over 2m} - 4 Z_h(k) {p_0^2\over 2m}\cr
p_h^{++}(k) &= {\kk^2\over 2m} + 4[B_h(k) k_0^2{2m\over p_0^2} +
A_h(k) {\kk^2\over 2m}] \virg v_0={p_0\over m} \cr } \Eq(3.43)$$
By using \equ(3.37), \equ(3.38) and \equ(2.15), it is easy to see that,
on the support of $T_0(k)$:
$$\eqalign{
 Z_h(k) = \bar Z_h(\g^h k) &=
 \sum_{h'=h}^0  z_{h'} t_0(\g^{-(h'-h)}k) =
 \sum_{h'=h+1}^0  z_{h'} +  z_h t_0(k) \cr
 A_h(k) = \bar A_h(\g^h k) &=
\sum_{h'=h}^0  \a_{h'} t_0(\g^{-(h'-h)}k) =
\sum_{h'=h+1}^0  \a_{h'} +  \a_h t_0(k) \cr
 B_h(k) = \bar B_h(\g^h k) &=
\sum_{h'=h}^0  \z_{h'} t_0(\g^{-(h'-h)}k) =
\sum_{h'=h+1}^0  \z_{h'} +  \z_h t_0(k) \cr
 E_h(k) = \bar E_h(\g^h k) &=
1+ 2\sum_{h'=h}^0  d_{h'} t_0(\g^{-(h'-h)}k) =
1+ 2\sum_{h'=h+1}^0  d_{h'} + 2 d_h t_0(k) \cr
 z_0 &=\e \quad,\quad \a_0 = \z_0 =  d_0 = 0 \cr
} \Eq(3.44)$$

Note that there are some relations between $\l_h$, $\m_h$ and $Z_h(0)$,
which can be defined in another equivalent way by the following steps.

\item{a)} We first integrate in a single step the fluctuations up to scale
$h+1$, without any free measure renormalization and by using the original
representation \equ(2.12) of the potential. The result of this operation can
be written in the form:
$$\tilde U_h(\f) = \sum_{s=1}^\io \int d\undx d\undy W_{h,s}(\undx,\undy)
\f^+_{x_1} \cdots \f^+_{x_s} \f^-_{y_1} \cdots \f^-_{y_s} \Eq(3.45)$$
where $\undx =(x_1,\ldots,x_s)$, $\undy =(y_1,\ldots,y_s)$. $U_h(\f)$
satisfies the identity:
$$\int P(d\psi) e^{-V(\f)} = {1\over\NN} \int P_0^{(\le h)}(d\psi) e^{-\tilde
U_h(\f)} \Eq(3.46)$$
where $P_0^{(\le h)}(d\psi)$ is the measure with propagator:
$$C_0^{(\le h)}(x)= \fra1{(2\p)^4}\ig dk\, t_0(\g^{-h}k) \fra{e^{-ikx}}{-i
k_0+\kk^2} \Eq(3.47)$$

\item{b)} We insert the representation \equ(2.16) of $\f^\pm$ in
\equ(3.45), we collect the terms containing the monomials
$(\prod_{i=1}^{r_1}\psi^+_{x_i}) (\prod_{j=1}^{r_2}\psi^-_{y_j})$,
with $2\le r_1+r_2 \le 4$, and we localize them, by a Taylor expansion of
the fields of order $0$, if $r_1+r_2>2$, and of order $2$, if $r_1+r_2=2$. We
obtain various local terms; those which do not contain any derivative of a
field are of the form
$$\l_h^{r_1,r_2} \int dx (\psi^+_x)^{r_1} (\psi^-_x)^{r_2} \Eq(3.48)$$
with
$$\l_h^{r_1,r_2} = \sum_{s=1}^\io {s\choose r_1} {s\choose r_2} \bar W_{h,s}
\r^{s-2} \virg \bar W_{h,s} = \lim_{\b|\L| \to \io} {1\over \b\L} \int d\undx
d\undy W_{h,s}(\undx,\undy) \Eq(3.49)$$

\item{c)} We insert the representation \equ(3.7) in \equ(3.48) and we
collect the terms proportional to $F_{40}$, $F_{21}$, $F_{02}-F_{20}$,
$F_{02}$, $D_{tt}$, $D_{ss}$ and $D_t$; let us call $\tilde \l_h$,
$\tilde \m_h$, $\tilde \n_h$, $2\tilde Z_h$, $2\tilde B_h$, $2\tilde A_h$ and
$(\tilde E_h-1)$ their respective coefficients.
By using \equ(3.49), it is easy to prove that:
$$\eqalign{
\tilde\l_h &= {1\over 4} \sum_{s=2}^\io \bar W_{h,s} \r^s \sum_{r=0}^4
{s\choose r} {s\choose 4-r} \cr
\tilde\m_h &= -{1\over \sqrt{2}} \sum_{s=2}^\io s(s-1)\bar W_{h,s} \r^s
\cr
\tilde Z_h &= {1\over 2} \sum_{s=2}^\io s(s-1) \bar W_{h,s} \r^s \cr}
\Eq(3.50)$$
Note that there is no contribution with $s=1$ in the equation for $\tilde
Z_h$, because they could come only from the monomials $\psi^+_x \psi^-_y$,
which only contribute to $F_{02}-F_{20}$, see \equ(3.19).

\item{d)} We absorb in the free measure the local terms proportional to
$F_{02}$, $D_{tt}$, $D_{ss}$ and $D_t$. If we call $U_h(\c)$ the remaining
part of
$\tilde U_h(\psi)$, thought as a function of $\c$, and $P_{0,B}^{(\le h)}(\c)$
the renormalized free measure, we can write:
$$\int P^{(\le 0)}(d\psi) e^{-V(\f)} = {1\over \NN}\int P_{0,B}^{(\le h)}(d\c)
e^{- U_h(\c)} \Eq(3.51)$$
However, this identity and the property that the local terms proportional to
$F_{02}$, $D_{tt}$, $D_{ss}$ and $D_t$ are equal to zero uniquely determine
the
effective potential $\\V_h(\c)$ and the measure $P_B^{(\le h)}(d\c)$. Hence
$\tilde\l_h = (p_0^2\r/2m) \l_h$, $\tilde\m_h = (p_0^2\r/2m)\m_h$ and (by
using the results of appendix 1) $\tilde Z_h t_0(\g^{-h}k) =
(p_0^2\r/2m)Z_h(k)$.

The last observation and \equ(3.50) imply the {\it exact} identity:
$$Z_h(0) = -{\m_h\over \sqrt{2}}  \virg \forall h\le 0 \Eq(3.52)$$
which will play a very important role in the discussion of the following
section.

\vskip2truecm

\vskip2.truecm
{\it \S4 The beta function.}
\vskip0.5truecm
\numsec=4\numfor=1\pgn=1

We want to study perturbatively the flow of the {\it running couplings}
$\l_h$, $\m_h$ and $\n_h$ (the {\it beta function} of our problem) and of the
{\it renormalization functions} $Z_h(k)$, $A_h(k)$, $B_h(k)$ and $E_h(k)$, by
keeping only the leading terms in the expansion of $r_{h-1}$ in
terms of $\{r_{h'}, h'\ge h\}$, if $r_h \equiv \{ \l_h, \m_h, \n_h, Z_h,
A_h, B_h, E_h \}$.

In order to understand which are the leading terms, we shall consider only the
Feynman graphs calculated by using the single scale propagator \equ(3.39);
this corresponds to the approximation of neglecting the irrelevant part of
$V_h(\c)$ in calculating the relevant part of $V_{h-1}(\c)$. In terms of the
tree expansion described in [G], this approximation implies that we consider
only trees with one tree vertex on scale $h$. The tree expansion allows to
prove also that one obtains in this way the same bounds that one would obtain
by summing over all trees with the same number of $\l$-vertices,
$\m$-vertices and $\n$-vertices on different scales.

In order to perform this analysis, we need bounds on the rescaled propagators
\equ(3.40). If we define a scale $h_0$ so that:
$$v_0^2 z_0  = \g^{2h_0} {q_0^2\over 4m^2}\Eq(4.1)$$
then we have to distinguish the case $h\ge h_0$ from the case $h\le h_0$. In
the first case, if $T_0(k) \not= 0$, $\g^{2h}\kk^4$ is dominating over
$a_h\kk^2$, at least
if $|a_h\kk^2| \le c z_0$, that we shall prove is true for $\l$ small enough.
In the second case $a_h\kk^2$ is dominating over $\g^{2h}\kk^4$.

If $\g^{2h} \ge Z_h(0)$, which essentially means $h\ge h_0$, and
$$|A_h| \le \eta \virg \g^{2h} |B_h| \le \eta \virg |E_h-1| \le
\eta \Eq(4.2)$$
where $\eta$ is some number sufficiently small, one can show that there exists
a fast decaying function $f(x)$ such that: %
$$|g_h^{\s_1\s_2}(x)| \le \g^{-h {\s_1+\s_2\over 2}} f(\g^h x_0, \xx)
\Eq(4.3)$$

Let us now consider a graph contributing to the flow equation for $\l_h$,
$\m_h$ or $Z_h(0)$ and denote $n_\s$ the number of external lines of type
$\c^\s$, $n_\l$ the number of $\l$-vertices, $n_\m$ the number of
$\m$-vertices.  In order to estimate the contribution of this graph,
we observe that the bound \equ(4.3) implies that each
internal half-line of type $\c^\s$ gives a contribution $\g^{-h\s/2}$, while
each one of the $(n_\l+n_\m-1)$ integrations that one has to perform to
evaluate the local part of the graph gives a contribution proportional to
$\g^{-h}$. Hence, up to a constant, the graph can be bounded by:
$$\eqalign{
&|\l_h|^{n_\l} |\m_h|^{n_\m} \g^{ h [ {1\over 2} (4n_\l +2n_\m-n_-)-
{1\over 2} (n_\m-n_+) -(n_\l+n_\m-1) ] } =\cr
&\qquad = \l_h|^{n_\l} |\m_h|^{n_\m} \g^{ h [n_\l - {1\over 2} n_\m
+{1\over 2} (n_+-n_-+2) ] } \cr} \Eq(4.4)$$

If we denote $N_L$ the number of independent loops of the graph (equal to the
number of propagators minus $(n_\l+n_\m-1)$) and we use the identity
\equ(3.52), we can write the bound \equ(4.4) in the form:
$$\l_h^{n_\l} (Z_h^2 \g^{-2h})^{n_\m/2} \g^{h(N_L+n_+)} \le
\l_h^{n_\l} Z_h^{n_\m/2} \g^{h(N_L+n_+)}
\Eq(4.5)$$

Note that the contribution of a graph to the beta function is obtained by
calculating its value at zero momentum of the external lines. Therefore no
graph with $N_L=0$ can contribute, because the single scale propagator
vanishes at zero momentum. Moreover, \equ(3.23) and \equ(3.38) imply
that $\l_0=Z_0/(4)$, so that,
by using \equ(4.5), one can easily show that, if $\l_0$ is small enough,:
$$Z_h \le 5 \r \l_h \Eq(4.6)$$
In fact this inequality follows by induction from the remark that, if it is
true for $h'\ge h$, then \equ(4.5) implies that the leading contributions
to the beta function are those with $N_L=1$. Hence one expects that there
exists a constant $c$ such that:
$$\eqalign{
|\l_{h-1}-\l_h| &\le c \l_h^2 \g^h \cr
|Z_{h-1}-Z_h| &= {|\m_{h-1}-\m_h|\over \sqrt{2}} \le {c\over \sqrt{2}}
\l_h^{3/2} \g^{2h} \cr } \Eq(4.7)$$
which implies \equ(4.6), together with
$$0<Z_h \le 2Z_0 \virg \hbox{if}\ 2h \ge h_0 \Eq(4.8)$$
if $\l_0$ is small enough.

Note that all the previous bounds rest on the hypothesis \equ(4.2); hence we
have to check that \equ(4.2) is consistent with \equ(4.6) and
\equ(4.8). This can be easily done, by noticing that the graphs
contributing to $B_{h-1}$ and $A_{h-1}$ are the graphs with $n_-=2$
and $n_+=0$ and that their contribution is of the form $\int dx x_0^2 W_G(x)$
and $\int dx \xx^2 W_G(x)$, respectively. Hence, we can use \equ(4.5) for
bounding $A_{h-1}- A_h$, while we have to multiply that bound by a
factor $\g^{-2h}$, in the case of $B_{h-1} -B_h$; by using also
\equ(4.6), we find that:
$$|B_{h-1} -B_h| \le c\l_h \g^{-h} \virg
|A_{h-1} -A_h| \le c\l_h \g^h \Eq(4.9)$$
In a similar way one can prove that
$$|E_{h-1}-E_h| = 2|d_h| \le c\l_h \g^h \Eq(4.10)$$
Hence \equ(4.2) are satisfied, if $\l_0$ is small enough.

\vskip.5truecm
{\bf Remark} - In order to make rigorous the previous considerations, one
can not really use a complete perturbative argument, because the formal series
representing the beta function are at most asymptotic series. In fact, by
applying the technique of reference [G], one can prove only $n!$ bounds on
the sum of all graphs with $n$ point vertices, like in the Fermi gas problem,
see [BG].
\vskip.5truecm

The previous discussion shows that, in order to study the beta function, it is
sufficient to consider carefully only the region $h <h_0$. The latter is the
region where, if, for some constants $c,c_1$:
$$\fra12 \le 1+4A_h \le c,\qquad cZ_h \le E_h^2+16 B_h Z_h \le c_1 Z_h,\quad
0<Z_h<2Z_0\=2\e\Eq(4.11)$$
the rescaled propagators $g_h^{--}(x)$ and $g_h^{-+}(x)$ are
essentially independent of $h$, that is they can be bounded by a rapidly
decaying function $f(x)$, uniformly in $A_h$, $B_h$, $Z_h$, $E_h$ verifying
\equ(4.11) (see appendix 2). On the contrary, the propagator
$g_h^{++}(x)$ can be bounded by $f(x)/Z_h$.

The previous properties of the rescaled propagators and the identity
\equ(3.52) imply that, in the bound of a generic graph contributing to the
beta function, two $\m_h$ vertices are essentially equivalent to one $\l_h$
vertex. A further simple analysis allows to prove that, if one wants to keep
in each flow equation only the leading terms, then one has to consider only
one loop graphs without $\n_h$ vertices.
Therefore, we shall study the flow equations, by keeping
only these contributions; the properties of the corresponding solutions will
be used to justify the approximation.

In this approximation $\n_h$ does not appear in the flow equations of $\l_h$
and $\m_h$; hence the growth of $\n_h$ is controlled
in a simple way by the right choice of $\n_0$, that is the right value of the
chemical potential, if one can control the flow of $\l_h$ and $\m_h$, which
only depend on the renormalization functions.

It turns out that the leading terms of the beta function depend in a smooth
way on the cutoff function $t_0(k)$, as one expects, because these terms
determine the asymptotic behaviour of the model, which is independent of the
choice of $t_0(k)$.  Therefore we shall approximate the smooth cutoff function
$t_0(k)$, whenever this is possible, by the characteristic function of the set
$\{k_0^2+\kk^2 \le 1\}$.

This approximation simplifies the discussion of the beta function, because it
implies that the supports of the Fourier transforms of the single scale
covariances $\bar C_h^{\s_1\s_2}(x)$ are disjoint.  This will guarantee that
the leading terms in the beta function for the running couplings on scale
$(h-1)$ will depend only on the running couplings on scale $h$; in fact we
have to consider only graphs with up to four point vertices, one loop and all
external momenta equal to $0$, so that all the propagators belonging to the
loop must have the same momentum.  In terms of the tree expansion described in
[G] this means that we must consider only trees with one tree vertex on scale
$h$.

Note that this approximation is apparently not allowed everywhere; in fact the
constants $\z_h$ and $\a_h$ (see \equ(3.33)) depend on $t_0'(k)^2$,
hence they are divergent when the regularization of $t_0(k)$ goes to zero.
However there is a cancellation of these contributions, as well as of all
contributions containing derivatives of $t_0(k)$, in the calculation of the
leading terms, as explained in appendix 2. Therefore also in this case we can
take the same approximation for the cutoff function.

There is another simplification following from the choice of $t_0(k)$
described before: $Z_h(k)$, $A_h(k)$ and $B_h(k)$ are independent of $k$, for
$k^2\le \g^h$, that is for the values of $k$ in the support of the Fourier
transform of $C_{\le h}^{\s_1\s_2}(x)$. This immediately follows from
\equ(3.37). Hence, also $Z_h$, $A_h$ and $B_h$ are constants.

There are very "few" graphs with one loop; therefore the calculation is
straightforward: but quite long.  Here we report the result, while the
details are exposed in appendix 2.  Neglecting terms of order
$\g^{2h}$ (which essentially come from the corrections to the scaling
of the propagators, which become quickly scale independent),
one finds after using \equ(3.52) to eliminate $\m_h$ from the row result
(see appendix 2):
$$\eqalignno{
\l_{h-1} &= \l_h - 36 q_0^3 \b_{2,h} (4Z_h)^2 \big( \l_h -
\fra{ Z_h}{6} \big)^2\cr
Z_{h-1} &= Z_h- \fra14  q_0^3\b_{2,h}(4Z_h)^3 Z_h \cr
E_{h-1} &=E_h - \fra14  q_0^3\b_{2,h}(4Z_h)^3 E_h &\eq(4.12)\cr
A_{h-1} &= A_h \cr
B_{h-1} &=B_h + \fra1{16} q_0^3\b_{2,h} (4Z_h)^2 E_h^2\cr
\n_{h-1}&= \g^2 \,\big(\n_h + q_0 \b_{1,h} (24Z_h\l_h
-4 Z_h^2)\big)\cr}$$
where:
$$\eqalign{
\b_{2,h} =& \fra{\log\g}{8\p^2 \sqrt{a_h^3 b_h}} \left(\fra2{v_0}\right)^3,
\quad \b_{1,h} = \fra{1-\g^{-2}}{8\p^2 ( \sqrt{a_h b_h} +a_h)}
q_0^2 \left(\fra2{v_0}\right)^3, \quad v_0={p_0\over m} \cr
a_h=&4Z_h(1+4 A_h),\quad b_h=(E_h^2+ 16Z_hB_h), \quad
q_0 = \fra{p_0^2}{2m} \cr}\Eq(4.13)$$

It is very easy to analyze this flow, under the conditions \equ(4.11),
implying that $\b_{2,h}$ is of order $Z_h^{-2}$. In fact, this observation is
sufficient to prove that $Z_h=O(1/|h|)$ for $h\to-\io$. But this property
has to be true also for $E_h$, since, by \equ(4.12):
$$\fra{E_{h-1}-E_h}{E_h} = \fra{Z_{h-1}-Z_h}{Z_h} \Eq(4.14)$$
It is now very easy to check that
$$B_h\tende{h\to-\io} B_{-\io}>0 \Eq(4.15)$$
while $A_h$ stays constant (indeed a small constant, roughly equal to its
value on scale $h_0$).

Finally, if we define:
$$c_0= \fra{ p_0^3 \log\g}{2\p^2 v_0^3 \sqrt{B_{-\io} (1+4A_{-\io})}}
\Eq(4.16)$$
we see that the first two equations
in \equ(4.12) can be written, for $h\to-\io$, in the form:
$$\eqalign{
\l_{h-1}=&\l_h- 36 c_0 Z_h^2 \left(\fra{\l_h}{Z_h} -\fra{1}{6}\right)^2\cr
Z_{h-1}=&Z_h- c_0  Z_h^2\cr }\Eq(4.17)$$

The discussion of the above equations is elementary and,
starting from initial data $Z_0=\e,\l_0=Z_0/4$ (or any other
close to them),  the result is that, if $\n_0$ is chosen so that $\n_h$ is
bounded uniformly in $h$, then, asymptotically:
$$Z_h= c\, |h|^{-1},\quad \l_h=\fra14 Z_h,\quad \n_h= O(\l_h) \Eq(4.18)$$
if $c$ is a suitable constant ($\e$ independent).

Note that these results are consistent with \equ(4.11), which can be then
proved inductively, together with \equ(4.18).

At this point it is very easy to check that all the neglected terms in the
beta function are at least of order $1/|h|^2$. Hence they can not change in a
substantial way the asymptotic properties of the flow (up to convergence
problems, see the remark above); only the values of $A_{-\io}$, $B_{-\io}$ and
$c$ depend on them, and $A_{-\io}$ has to be a small number (of order $\e$).
Note that this last observation is important, in order to be sure that
the first condition in \equ(4.11) is preserved, since we do not have a
control on the sign of $A_{-\io}$.

The main consequence of the previous discussion is that, for $k\to 0$ (that is
for $h\to-\io$), the model is gaussian (asymptotic freedom) and the pair
Schwinger function of the fields $\psi^\pm$ behaves as:
$$\tilde S_{-+}(k) = -\tilde S_{--}(k)= -\tilde S_{++}(k) \simeq
{ q_0\over 8B_{-\io}} {1\over k_0^2 + c^2 \kk^2} \Eq(4.19)$$
where the sound speed $c$ is given by:
$$c^2 = {(1+4A_{-\io}) v_0^2\over 16 B_{-\io}} \Eq(4.20)$$

\vskip2truecm

\vglue2.truecm
{\it Appendix 1: Propagators for the Bose gas.}
\vglue0.5truecm\numsec=1\numfor=1\pgn=1

Let $P^{(t)}(d\psi)$ be the formal complex gaussian measure with covariance

$$g^{(t)}(x) = {1\over (2\p)^4} \ig dk\, e^{-ikx}
\fra{t(k)}{-i k_0+ {\kk^2\over 2m} }\Eqa(A1.1)$$
where $x=(x_0,\xx)$, $k=(k_0,\kk)$ and $t(k)$ is a positive cutoff function.

We consider the fields $\chi^\pm_x=\fra1{\sqrt{2\r}}(\psi_x^+\pm\psi_x^-)$.
Their propagator has the form:
$$\media{\chi^\s_x\chi^{\s'}_y}=\fra1{(2\p)^4}\ig e^{-ikx} t(k)
G^{-1}(k)_{\s\s'}\Eqa(A1.2)$$
where the matrix $G$, which we call the propagator matrix, is:
$$G=\r\pmatrix{\fra{\kk^2}{2m}&ik_0\cr
-ik_0&-\fra{\kk^2}{2m}\cr}\Eqa(A1.3)$$
Therefore the formal gaussian measure:
$$P^{(t)}(d\psi)
e^{-\r\sum_{\s\s'}\ig \chi^\s_{\s k}  \D_{\s\s'} \chi^{\s'}_{-\s' k}
dk}\Eqa(A1.4)$$
has a propagator matrix $G'\= G+2\r \D t(k)$. In particular, if $\D$
is associated with the quadratic form:
$$\ig dx [2a (\chi^+_x)^2 - 2 \sum_{i=0}^3
b_i(\dpr_i\chi^-_x)^2 -2c \chi^+_x \dpr_0 \chi^-_x]\Eqa(A1.5)$$
with $a,b_0,b=b_1=b_2=b_3, c$ not negative real numbers, that is:
$$\D= \pmatrix{2a & ick_0 \cr-ick_0 & -2(b_0 k_0^2+ b\kk^2)\cr}\Eqa(A1.6)$$
then it is:
$$G'=\r\pmatrix{\fra{\kk^2}{2m}+4at(k) & i k_0 [1+2ct(k)] \cr
-ik_0[1+2ct(k)] &-\fra{\kk^2}{2m} -4 (b k_0^2+ b\kk^2) t(k)\cr}\Eqa(A1.7)$$
And the $\psi^\pm$ fields propagators with respect to the measure
$P^{(t)}_{ab}(d\psi)$ with propagator matrix $G'$ can be immediately
checked to be:
$$\eqalign{
&\qquad \ig P^{(t)}_{a,b}(d\psi) \psi^-_x\psi^+_y =
\fra{1}{(2\p)^4} \ig dk\,
e^{-ik(x-y)}\cdot\cr
&\cdot \fra{ik_0[1+2ct(k)] +{\kk^2\over 2m} +2d^+(k) t(k)}{ [1+2ct(k)]^2 k_0^2
+{\kk^4\over 4m^2} +4d^+(k) {\kk^2\over 2m} t(k) +16a(\sum_{i=0}^3
b_ik_i^2)t(k)^2} \,t(k) \cr } \Eqa(A1.8)$$
and:
$$\eqalign{
&\qquad \ig P^{(t)}_{a,b}(d\psi) \psi^+_x\psi^+_y = \ig P^{(t)}_{a,b}
(d\psi) \psi^-_x\psi^-_y = \fra{1}{(2\p)^4} \ig dk\, e^{-ik(x-y)}\cdot
\cr&\cdot \fra{-2d^-(k) t(k)}{ [1+2ct(k)]^2 k_0^2
+{\kk^4\over 4m^2} +4d^+(k) {\kk^2\over 2m} t(k) +16a(\sum_{i=0}^3
b_ik_i^2)t(k)^2} \,t(k) \cr} \Eqa(A1.9)$$
where $d^\pm(k) =a \pm (b_0 k_0^2 + b\kk^2)$. Note that \equ(A1.9)
can be derived also by performing a Bogoliubov transformation.

The previous calculation can be immediately generalized, in the sense that, if
the measure $P^{(t)}_G(d\psi)$ has propagator matrix $G$ and cutoff function
$t(k)$, then the measure
$$P^{(t)}_G(d\psi)
e^{-\r\sum_{\s\s'}\ig \chi^\s_{\s k}  \D_{\s\s'} \chi^{\s'}_{-\s' k}
dk}\Eqa(A1.10)$$
has propagator matrix
$$G' = G+2\r t(k)\D \Eqa(A1.11)$$

\vskip2truecm

\vglue2.truecm
{\it Appendix 2: The second order beta function.}
\vglue0.5truecm\numsec=2\numfor=1\pgn=1

In this appendix we want to prove \equ(4.12). We first note that all the
terms in the r.h.s. of \equ(4.12) are obtained by applying the localization
operator to:
$$V_h^{\le 4} \= \EE_h(V_h) - {1\over 2!} \EE^T_h(V_h^2) + {1\over 3!}
\EE^T_h(V_h^3) - {1\over 4!} \EE^T_h(V_h^4) \Eqa(A2.1)$$
where $\EE_h$ and $\EE^T_h$ denote, respectively, the expectation and the
truncated expectation with respect to the measure describing the fluctuations
on scale $h$, whose propagator is given by \equ(3.39)-\equ(3.43).

Some other remarks are important.

\noindent 1) In order to calculate the beta function one has to evaluate some
Feynman graphs at zero momentum of the external lines. Therefore only terms
without internal lines or terms with at least one loop can contribute, since
the single scale propagator vanishes at zero momentum.

\noindent 2) The previous remark also implies that, in the graphs with only
one loop, all internal lines must carry the same momentum.  Hence, if we
suitably choose the cutoff function $t_0(k)$, the internal lines of
the loop may only have propagators of scale $h$ or $h+1$; in fact at least one
propagator must be of scale $h$ and the supports of the Fourier transforms of
the propagators of scale $h$ and $h'\ge h$ are disjoint if $h'>h+1$.  This
implies that the trees (see [G] for the definitions), that one has to consider
in evaluating the contribution of such graphs, are the trees with only one
vertex of scale $h$ and at most three endpoints and the trees with one vertex
on scale $h$ and one non trivial vertex on scale $h+1$, corresponding to some
irrelevant contribution.

\noindent 3) In the calculation of the graphs with only one loop, the subgraph
associated with the tree vertex of scale $h+1$ has no loop. Therefore in this
tree vertex the $\RR$ operation coincides with the identity.

\noindent 4) Since we are interested only in the leading orders, we can
neglect in the rescaled propagators \equ(3.40) the terms proportional to
$\g^{2h}$ and the dependence on $k$ of $Z_h$, $A_h$, $B_h$ and $E_h$
(see \equ(3.44)).  For the same reason we can approximate $Z_{h+1}$,
$A_{h+1}$, $B_{h+1}$ and $E_{h+1}$ by $Z_h$, $A_h$, $B_h$ and $E_h$ in
the expression of $g_{h+1}^{\s_1\s_2}(x)$.  Finally we can neglect the
difference between $\l_{h+1}$, $\m_{h+1}$ and $\l_h$, $\m_h$ in the
endpoints of the trees involving a tree vertex on scale $h+1$.

The previous remarks imply that the leading terms in the beta function can be
obtained by the following steps:
\*
\0{a)} Evaluate the graphs with one loop and propagator given by
the sum of the single scale propagators of scale $h$ and $h+1$,
approximated as explained in remark 4), that is:
$$\g^{(3+\fra12(\s_1+\s_2))h}g_{1,h}^{\s_1\s_2}(\g^h x) \Eqa(A2.2)$$
where
$$g^{\s_1\s_2}_{1,h}(x) = {1\over (2\p)^4} \int dk \ e^{-ikx}
{T_1(k) \, \bar p_h^{\s_1\s_2}(k) \over \r\DD_h(k)} \Eqa(A2.3)$$
Here $\DD_h(k)$ and $\bar p_h^{\s_1\s_2}$ denote the expressions \equ(3.41)
and \equ(3.43), modified as explained in remark 4) above, that is:
$$\eqalign{
&\DD_h(k) = b_h k_0^2 + a_h {v_0^2\over 4}\kk^2 \cr
&a_h = 4Z_h (1+4A_h) \virg b_h = E_h^2 + 16 Z_h B_h \cr}
\Eqa(A2.4)$$

$$\eqalign{
\bar p_h^{--}(k) &= -4q_0 Z_h \virg \bar p_h^{-+}(k)= \bar
p_h^{+-}(-k) = -ik_0E_h \cr
\bar p_h^{++}(k) &= {16Z_h B_h k_0^2 + a_h\kk^2
{v_0^2\over 4} \over 4q_0 Z_h} \virg q_0={p_0^2\over 2m} ,\qquad v_0={p_0\over
m} \cr} \Eqa(A2.5)$$
and
$$T_1(k) = T_0(k) + T_0(\g^{-1}k) = t_0(\g^{-1}k) - t_0(\g k) \Eqa(A2.6)$$
\*
\0{b)} Evaluate the same graphs with propagator of scale $h+1$, again
approximated as in remark 4). This propagator is obtained from \equ(A2.3) by
substituting $T_1(k)$ with
$$T_2(k) = T_0(\g^{-1}k) = t_0(\g^{-1}k) - t_0(k) \Eqa(A2.7)$$
We shall denote $g^{\s_1\s_2}_{2,h}$ the corresponding rescaled propagator.
\*
\0{c)} Subtract the values found in b) from the values found in a) and
add the trivial graphs without any internal line.
\*
\0{d)} Approximate in the result the cutoff function $t_0(k)$ by the
characteristic function of the set
$\{k_0^2 +{\kk^2\over 2m} {p_0^2\over 2m} \le ({p_0^2\over 2m})^2\}$.
Note that this
approximation is everywhere equivalent to calculating the graphs with all
propagators on the single scale $h$, except in the case of the beta function
for $B_{h-1}$, $A_{h-1}$ and $E_{h-1}$, which involve derivatives with respect
to the loop momentum.  Hence, except in this case, we have to calculate the
graphs by using the propagator obtained from \equ(A2.3) by substituting
$T_1(k)$ with $T_0(k)$; we shall denote again $g^{\s_1\s_2}_h$ this rescaled
propagator.
\*
The trivial graphs give the linear terms in the r.h.s.  of
\equ(4.12), except the term linear in $\l_h$ appearing in the equation for
$\n_{h-1}$.  This term is obtained by contracting two $\c^-$ fields in
the $\l_h$ vertex; one gets:
$$-{4\choose 2} {p_0^2\over 2m}\l_h \g^{2h} 4q_0 Z_h \b_{1,h} F_{20}
\Eqa(A2.8)$$
where $F_{20}$ is defined as in \equ(3.18) and
$$\b_{1,h} = \int {dk\over (2\p)^4} {T_0(k)\over \DD_h(k)} =
\fra{1-\g^{-2}}{8\p^2 ( \sqrt{a_h b_h} +a_h)} q_0^2
\left( \fra2{v_0}\right)^3 \Eqa(A2.9)$$

The quadratic terms in the equations for $\l_h$ and $\m_h$ are associated with
the graphs drawn in Fig. \equ(A2.10), where the heavy lines represent the
$\c^-$ fields and the dotted ones represent $\c^+$.

Note that the coefficients in front of the different graphs, here and
in the following figures, indicate how many times the graph appears, if
one expands the powers of the potential in the r.h.s.  of \equ(A2.1) in
terms of the different monomials of the field, whose sum gives the
potential, and consider the different possibilities of connecting
different point vertices, giving rise to the same graph.  In order to
get the right contribution to the beta function, one has to consider
also the coefficients of the expectations in \equ(A2.1) and the
combinatorial factors which count the different possibilities of
choosing the external lines in the different vertices of the
graph and the
different possibilities of contracting the internal lines.

\initfig{figa61}
\write13<30 30 moveto 50 50 lineto 30 70 lineto>
\write13<120 30 moveto 100 50 lineto 120 70 lineto>
\write13<50 50 moveto 75 65 lineto 100 50 lineto 75 35 lineto 50 50 lineto>
\write13<180 30 moveto 200 50 lineto 180 70 lineto>
\write13<200 50 moveto 225 65 lineto 250 50 lineto 225 35 lineto>
\write13<200 50 lineto>
\write13<250 50 280 50 tlinea>
\write13<stroke>
\endfig

\insertplot{300pt}{100pt}{%
\ins{140pt}{55pt}{$+\quad 2$}}{figa61}{\eqa(A2.10)}

The contribution to $\LL V_{h-1}(\c)$ coming from the graphs in
\equ(A2.10) is:
$$-{1\over 2} \, \Big[ {4\choose 2}^2 2 \l_h^2 (4q_0Z_h)^2 \b_{2,h} F_{40} +
2 {4\choose 2} 2 \l_h\m_h (4q_0Z_h)^2 \b_{2,h} F_{21} \Big]
\left({p_0^2\over 2m}\right)^2\Eqa(A2.11)$$
where
$$\b_{2,h} = \int {dk\over (2\p)^4} {T_0(k)\over \DD_h(k)^2}
= \fra{\log\g}{8\p^2 \sqrt{a_h^3 b_h}} \left(\fra2{v_0}\right)^3\Eqa(A2.12)$$

The cubic terms in the equations for $\l_h$ and $\m_h$ are associated
with the graphs drawn in \equ(A2.13).

\initfig{figa62}
\write13<0.8 0.8 scale 20 90 moveto 30 110 lineto 20 130 lineto>
\write13<30 110 moveto 60 90 lineto 80 80 lineto>
\write13<30 110 moveto 60 130 lineto 80 140 lineto>
\write13<60 90 60 130 tlinea>
\write13<110 90 moveto 120 110 lineto 110 130 lineto>
\write13<120 110 moveto 150 90 lineto 170 80 lineto>
\write13<120 110 moveto 135 120 lineto 135 120 150 130 tlinea>
\write13<170 140 moveto 150 130 lineto 150 110 lineto>
\write13<150 110 150 90 tlinea>
\write13<220 90 moveto 230 110 lineto 220 130 lineto>
\write13<230 110 moveto 245 100 lineto 245 100 260 90 tlinea>
\write13<230 110 moveto 245 120 lineto 245 120 260 130 tlinea>
\write13<280 80 moveto 260 90 lineto 260 130 lineto 280 140 lineto>
\write13<10 30 30 30 tlinea>
\write13<30 30 moveto 60 10 lineto 80 0 lineto>
\write13<60 10 60 50 tlinea>
\write13<30 30 moveto 60 50 lineto 80 60 lineto>
\write13<110 30 130 30 tlinea>
\write13<130 30 moveto 160 10 lineto 180 0 lineto>
\write13<130 30 moveto 145 40 lineto 145 40 160 50 tlinea>
\write13<180 60 moveto 160 50 lineto 160 30 lineto>
\write13<160 30 160 10 tlinea>
\write13<220 30 240 30 tlinea>
\write13<240 30 moveto 255 40 lineto 255 40 270 50 tlinea>
\write13<240 30 moveto 255 20 lineto 255 20 270 10 tlinea>
\write13<290 60 moveto 270 50 lineto 270 10 lineto 290 0 lineto>
\write13<stroke>
\endfig

\insertplot{240pt}{120pt}{%
\ins{4pt}{92pt}{$\st 3$}
\ins{68pt}{92pt}{$\st +\ 6$}
\ins{148pt}{92pt}{$\st +\ 3$}
\ins{232pt}{92pt}{$\st +$}
\ins{0pt}{28pt}{$\st 3$}
\ins{68pt}{28pt}{$\st +\ 6$}
\ins{152pt}{28pt}{$\st +\ 3$}
}{figa62}{\eqa(A2.13)}

The contribution to $\LL V_{h-1}(\c)$ coming from the graphs in \equ(A2.13)
is:
$${1\over 6} \, 3 \cdot 8 \Big[ {4\choose 2} \l_h\m_h^2 F_{40} + \m_h^3
F_{21} \Big] \left({p_0^2\over 2m}\right)^3 \b_{3,h} \Eqa(A2.14)$$
where:
$$\eqalign{
\b_{3,h} &= \r^3\int {dk\over (2\p)^4} \Big[ g^{--}_h (k)^2 g^{++}_h(k) +
2 g^{--}_h (k) g^{-+}_h(k)^2 +\cr
&+ g^{--}_h (k) g^{-+}_h(k) g^{+-}_h (k)
\Big] = \cr
&=\r^3 \int {dk\over (2\p)^4} \Big[ g^{--}_h (k)^2 g^{++}_h(k) +
g^{--}_h (k) g^{-+}_h(k)^2 \Big] = 4q_0Z_h \b_{2,h} \cr} \Eqa(A2.15)$$
The last equality follows from the identity:
$$\r^2 [g^{--}_h(k) g^{++}_h(k) + g^{-+}_h(k)^2] = -{T_0(k)^2\over \DD_h(k)}
\Eqa(A2.16)$$
and from the observation that $T_0(k)^2=T_0(k)$, in the approximation of item
d) above. We also used the fact that $g^{-+}_h(k) = - g^{+-}_h(k) =
-g^{-+}_h(-k)$. Hence \equ(A2.14) can be written as:
$$4 \Big[ {4\choose 2} \l_h\m_h^2 F_{40} + \m_h^3
F_{21} \Big] 4q_0Z_h \left({p_0^2\over 2m}\right)^3 \b_{2,h} \Eqa(A2.17)$$

The quartic term in the equation for $\l_h$ is associated with the graphs
drawn in \equ(A2.18).

\initfig{figa64}
\write13<10 130 moveto 90 130 lineto>
\write13<10 90 moveto 90 90 lineto>
\write13<30 130 30 90 tlinea>
\write13<70 130 70 90 tlinea>
\write13<110 130 moveto 190 130 lineto>
\write13<110 90 moveto 130 90 lineto 130 90 150 90 tlinea 150 90 moveto>
\write13<190 90 lineto>
\write13<130 130 130 110 tlinea 130 110 moveto 130 90 lineto>
\write13<170 130 170 90 tlinea>
\write13<210 130 moveto 250 130 lineto 250 130 270 130 tlinea 270 130 moveto>
\write13<290 130 lineto>
\write13<210 90 moveto 230 90 lineto 230 90 250 90 tlinea 250 90 moveto>
\write13<290 90 lineto>
\write13<230 130 230 110 tlinea 230 110 moveto 230 90 lineto>
\write13<270 130 moveto 270 110 lineto 270 110 270 90 tlinea>
\write13<40 60 moveto 120 60 lineto>
\write13<40 20 moveto 60 20 lineto 60 20 100 20 tlinea 100 20 moveto>
\write13<120 20 lineto>
\write13<60 60 60 40 tlinea 60 40 moveto 60 20 lineto>
\write13<100 60 100 40 tlinea 100 40 moveto 100 20 lineto>
\write13<180 60 moveto 260 60 lineto>
\write13<180 20 moveto 220 20 lineto 220 20 240 20 tlinea 240 20 moveto>
\write13<260 20 lineto>
\write13<200 60 200 20 tlinea>
\write13<240 60 240 40 tlinea 240 40 moveto 240 20 lineto>
\write13<stroke>
\endfig

\insertplot{300pt}{150pt}{%
\ins{10pt}{45pt}{$+\ 12$}
\ins{140pt}{45pt}{$+\ 12$}
\ins{90pt}{115pt}{$+\ 12$}
\ins{-5pt}{115pt}{$+\ 6$}
\ins{190pt}{115pt}{$+\ 6$}}{figa64}{\eqa(A2.18)}

The contribution to $\LL V_{h-1}(\c)$ coming from the graphs in \equ(A2.18)
is:
$$-{1\over 4!}\, 2^4 \cdot 6 \left({p_0^2\over 2m}\right)^4 \m_h^4 \b_{4,h}
F_{40} \Eqa(A2.19)$$
where:
$$\eqalign{
&\b_{4,h} = \r^4 \int {dk\over (2\p)^4} \Big[ g^{--}_h (k)^2 g^{++}_h(k)^2 +
2 g^{--}_h(k) g^{-+}_h(k)^2 g^{++}_h(k) + \cr
&+g^{-+}_h(k)^4 - 2 g^{--}_h(k) g^{-+}_h(k)^2 g^{++}_h(k)
+ 2 g^{--}_h(k) g^{-+}_h(k)^2 g^{++}_h(k) \Big] = \cr
&=\r^4 \int {dk\over (2\p)^4} \Big[ g^{--}_h (k) g^{++}_h(k) +
g^{-+}_h(k)^2 \Big]^2 = \b_{2,h} \cr} \Eqa(A2.20)$$
where we used again the identity \equ(A2.16).

The leading contribution in the flow equation for $Z_h$ is the
quadratic term associated with the graph in the first line of
\equ(A2.22), whose contribution to the local part is:
$$-{1\over 2} \ 2 \left({p_0^2\over 2m}\right)^2 \m_h^2 (4q_0Z_h)^2 \b_{2,h}
F_{02} \Eqa(A2.21)$$

\initfig{figa63}
\write13<90 100 120 100 tlinea 120 100 180 100 arco 180 100 120 100 arco>
\write13<180 100 210 100 tlinea>
\write13<170 40 moveto 200 40 lineto 260 40 moveto 290 40 lineto>
\write13<260 40 moveto 230 20 lineto 200 40 lineto>
\write13<200 40 230 60 tlinea 230 60 260 40 tlinea>
\write13<20 40 moveto 50 40 lineto 110 40 moveto 140 40 lineto>
\write13<50 40 80 60 tlinea 80 60 moveto 110 40 lineto>
\write13<110 40 80 20 tlinea 80 20 moveto 50 40 lineto>
\write13<stroke>
\endfig

\insertplot{300pt}{140pt}{%
\ins{155pt}{45pt}{$+$}}{figa63}{\eqa(A2.22)}

\0and no operators involving field derivatives arise because this is a
marginal operator.

The contribution to the local part of the last two graphs of \equ(A2.22) is:
$$\eqalign{
&-{1\over 2} 2^2 \left({p_0^2\r\over 2m}\right)^2 \m_h^2 \Big\{ \g^{2h} F_{20}
\int {dk\over (2\p)^4} \Big[ g_h^{--}(k) g_h^{++}(k) + g_h^{-+}(k)^2 \Big]-\cr
&-   {1\over 2} [\b_{t,h}^{(1)} - \b_{t,h}^{(2)}] q_0^2 D_{tt} -
{1\over 6} [\b_{s,h}^{(1)} - \b_{s,h}^{(2)}] 4m^2 q_0^2 D_{ss} \Big\}
\cr}\Eqa(A2.23)$$
where $D_{tt}$ and $D_{ss}$ are defined as in \equ(3.20) and
$$\eqalign{
\b_{t,h}^{(i)} &= \int dx\ x_0^2 \Big[ g_{i,h}^{-+}(x)^2 - g_{i,h}^{--}(x)
g_{i,h}^{++}(x) \Big] \cr
\b_{s,h}^{(i)} &= \int dx\ \xx^2 \Big[ g_{i,h}^{-+}(x)^2 - g_{i,h}^{--}(x)
g_{i,h}^{++}(x) \Big] \cr } \Eqa(A2.24) $$

$$g^{\s_1\s_2}_{i,h}(x) = {1\over (2\p)^4} \int dk
\fra{e^{-ikx} \, T_i(k) \, \bar p_h^{\s_1\s_2}(k)}{\r\DD_h(k)} \Eqa(A2.25)$$

We have:
$$\eqalign{
\r^2\b_{t,h}^{(i)} &= \int {dk\over (2\p)^4} \Big\{ E_h^2
\Big[ {\dpr\over \dpr k_0} {k_0 T_i\over \DD_h} \Big]^2 +
\Big[ {\dpr\over \dpr k_0} {T_i\over \DD_h} \Big]
{\dpr\over \dpr k_0} \Big[ \Big( 1-
{E_h^2 k_0^2\over \DD_h} \Big) T_i \Big]
\Big\} \cr
&= \int {dk\over (2\p)^4} \Big\{ E_h^2{T_i^2\over \DD_h^2} +
\Big({\dpr \over \dpr k_0}{T_i\over \DD_h} \Big) {\dpr T_i\over \dpr k_0}
\Big\} \cr }\Eqa(A2.26)$$

It is easy to see that the terms containing the derivatives of $T_i$ give the
same contribution to $\b_{t,h}^{(1)}$ and $\b_{t,h}^{(2)}$, so that, by doing
in the remaining term the approximation of item d) above, we get:
$$\r^2 [\b_{t,h}^{(1)} - \b_{t,h}^{(2)}] = \b_{2,h} E_h^2 \Eqa(A2.27)$$

We have also:
$$\eqalign{
\r^2 \b_{s,h}^{(i)} &= \sum_{j=1}^3 \int {dk\over (2\p)^4} \Big\{ E_h^2
\Big[ {\dpr\over \dpr k_j} {k_0 T_i\over \DD_h} \Big]^2 +
\Big[ {\dpr\over \dpr k_j} {T_i\over \DD_h} \Big]
{\dpr\over \dpr k_j} \Big[ \Big( 1- {E_h^2 k_0^2\over \DD_h} \Big) T_i \Big]
\Big\} =\cr
&= \sum_{j=1}^3 \int {dk\over (2\p)^4}
\Big({\dpr \over \dpr k_j}{T_i\over \DD_h} \Big) {\dpr T_i\over \dpr k_j}
\cr }\Eqa(A2.28)$$
and one can see again that the terms containing the derivative of $T_i$ give
the same contribution to $\b_{s,h}^{(1)}$ and $\b_{s,h}^{(2)}$, so that:
$$\b_{s,h}^{(1)} - \b_{s,h}^{(2)} = 0 \Eqa(A2.29)$$

By summarizing, the contribution to the local part of the last two graphs of
\equ(A2.22) is:
$$2 \b_{1,h} \left({p_0^2\over 2m}\right)^2 \m_h^2 \g^{2h} F_{20} + \b_{2,h}
E_h^2 \left({p_0^2\over 2m}\right)^2 \m_h^2 q_0^2 D_{tt} \Eqa(A2.30)$$

\initfig{figa65}
\write13<90 40 120 40 tlinea 120 40 moveto 150 60 lineto 180 40 lineto>
\write13<210 40 lineto 120 40 moveto 150 20 lineto 150 20 180 40 tlinea>
\write13<stroke>
\endfig

\insertplot{300pt}{80pt}{}{figa65}{\eqa(A2.31)}

The leading contribution in the flow equation for $E_h$ is associated with the
graph drawn in \equ(A2.31), whose local part is calculated by taking the
first order Taylor expansion of the external $\chi^-$ field. The term of order
zero, which would contribute to the relevant term $F_{11}$, cancels out for
parity reasons, as well as the term of order one containing the spatial
derivatives, so that the local part is equal to:
$$-{1\over 2} \ 2 \ 2 \ 2 \left({p_0^2\over 2m}\right)^2 \m_h^2
(\b^{(1)}_{5,h} - \b^{(2)}_{5,h}) q_0 (-D_t) \Eqa(A2.32)$$
where $D_t$ is defined as in \equ(3.20) and, if we use also the definition
\equ(A2.25):
$$\b_{5,h}^{(i)} = \r^2 \int dx\ x_0 g_{i,h}^{--}(x) g_{i,h}^{+-}(x)
\Eqa(A2.33)$$

\0We have:
$$\b_{5,h}^{(i)} = 4 q_0 Z_h E_h \int {dk\over (2\p)^4} {k_0 T_i\over \DD_h}
\Big[ {\dpr\over \dpr k_0} {T_i\over \DD_h} \Big] \Eqa(A2.34)$$
and we can see, as before, that the terms containing the derivatives of
$T_i$ give the same contribution to $\b_{5,h}^{(1)}$ and $\b_{5,h}^{(2)}$,
so that, in the usual approximation:
$$\b_{5,h}^{(1)} - \b_{5,h}^{(2)} = - 8 q_0 Z_h E_h b_h \int {dk\over (2\p)^4}
{k_0^2 T_0(k) \over \DD_h^3} = -2 q_0 Z_h E_h \b_{2,h} \Eqa(A2.35)$$
where the last equality follows from an explicit calculation and $\b_{2,h}$ is
exactly the same function of $a_h$ and $b_h$ defined in \equ(A2.12). Hence the
local part of the graph drawn in \equ(A2.31) can be written as:
$$-8 \left({p_0^2\over 2m}\right)^2 q_0^2 \m_h^2 Z_h E_h \b_{2,h} D_t
\Eqa(A2.36)$$

The flow equations immediately follow from \equ(A2.8),
\equ(A2.11), \equ(A2.17), \equ(A2.19), \equ(A2.21), \equ(A2.30) and
\equ(A2.36):
$$\eqalignno{
\l_{h-1} &= \l_h -36 q_0^3 (4 Z_h)^2 \b_{2,h} \left[ \l_h -
\fra{\m_h^2}{12 Z_h} \right]^2 \cr
\m_{h-1} &= \m_h -12 q_0^3 (4 Z_h)^2 \b_{2,h} \m_h \left[ \l_h -
\fra{\m_h^2}{12 Z_h} \right] \cr
Z_{h-1}  &= Z_h - 8 q_0^3 Z_h^2 \b_{2,h} \m_h^2
&\eqa(A2.37)\cr
A_{h-1}  &= A_h \cr
B_{h-1}  &= B_h + {1\over 2}  q_0^3\b_{2,h} E_h^2 \m_h^2 \cr
E_{h-1}  &= E_h -8 q_0^3Z_h E_h \b_{2,h} \m_h^2\cr
\n_{h-1} &= \g^2 \Big[ \n_h + 24 q_0Z_h \b_{1,h} \l_h - 2  q_0\b_{1,h}
\m_h^2 \Big]\cr }$$
where:
$$\b_{2,h} = \fra{\log\g}{8\p^2 \sqrt{a_h^3 b_h}}
\left(\fra2{v_0}\right)^3 \quad,\quad
\b_{1,h} = \fra{1-\g^{-2}}{8\p^2 ( \sqrt{a_h b_h} +a_h)}
q_0^2 \left(\fra2{v_0}\right)^3\Eqa(A2.38)$$
with:
$$a_h=4 Z_h(1+4A_h) \qquad,\qquad b_h=E_h^2 +16 Z_h B_h \Eqa(A2.39)$$
and the \equ(4.12) follow immediately from the above by replacing
$\m_h$ via \equ(3.52).

One may wonder whether it might be that the equation for $\m_h$ is
compatible with the others. Of course, if \equ(3.52) is valid, it must be;
this can also be seen directly (and in fact it is this remark that would
make one conjecture the exact relation \equ(3.52), if one did not know
it). However the compatibility is true only to the order of the
calculation that we are performing, \ie within the neglected
corrections: if we did not know \ap the validity of \equ(3.52) we could
not guarantee that the corrections and the incertitude on the initial
data would not spoil the relation \equ(3.52) and turn the flow equation
into an unmanageable relation of little use.

\vskip2truecm

\vglue2.truecm
{\it References.}
\pgn=1\numsec=0

\vglue0.5truecm
\halign{
\\[#]\hfill&\vtop{\hsize=15.truecm\\#}\cr
\noalign{\vskip2.truept}
ADG&{Abrikosov, A.A., Gorkov, L.P., Dzyaloshinski, I.E.: {\it Methods of
  quantum field theory in Statistical Physics}, Dover Publications,
  1963.}\cr\noalign{\vskip2.truept}
B&{Benfatto, G.: {\it Renormalization group approach to zero temperature
  Bose condensation problem}, work in progress,
  1994.}\cr\noalign{\vskip2.truept}
BGPS&{Benfatto, G., Gallavotti, G., Procacci, A., Scoppola, B.: {\it
  Beta function and Sch\-win\-ger functions for many fermion systems in one
  dimension. Anomaly of the Fermi surface},
  Communications in  Mathematical Physics, {\bf
  160}, 93--172, 1994.}\cr\noalign{\vskip2.truept}
BG&{Benfatto, G., Gallavotti, G.: {\it Perturbation theory of the Fermi
  surface in a quantum liquid. A general quasi particle formalism and one
  dimensional systems},  Journal of Statistical Physics {\bf 59},
  541--664, 1990}\cr\noalign{\vskip2.truept}
Bo&{Bogoliubov, N. N.: Journal of Physics (USSR) {\bf II}, 23,
  1947.}\cr\noalign{\vskip2.truept}
G&{Gallavotti, G.: {\it
  Renormalization theory and ultraviolet stability via
  re\-nor\-ma\-li\-za\-tion
  group methods}, Reviews of Modern Physics, {\bf 57}, 471--569, 1985.}
  \cr\noalign{\vskip2.truept}
Gi&{Ginibre, J.: {\it On the asymptotic exactness of the Bogoliubov
  approximation for many boson systems},
  Communications in Mathematical Phys\-ics {\bf 8},
  26--51, 1968.}\cr\noalign{\vskip2.truept}
GK&{Gawedzky, K., Kupiainen, A.:
  {\it Massless lattice $\phi^4_4$ theory: rigorous control of a
  renormalizable asymptotically free model}, Communications in Mathematical
  Phys\-ics, {\bf 99}, 197--252, 1985.} \cr\noalign{\vskip2.truept}
HP&{Hugenholtz, N.M., Pines, D.: {\it Ground-state energy and excitation
  spectrum of a system of interacting bosons}, Physical Review {\bf 116},
  489, 1959.}\cr\noalign{\vskip2.truept}
JMMS&{Jimbo, M, Miwa, T., Mori, Y., Sato, M.: {\it Density matrix of an
  impenetrable gas and the fifth Painlev\'e transcendent}, Physica {\bf D1},
  80--158, 1980.}\cr\noalign{\vskip2.truept}
KLS&{Kennedy, T., Lieb, E., Shastry: {\it The $XY$ model has long
  range order for all spins and all dimensions greater than one},
  Physical Review Letters, {\bf 61}, 2582--2584,
  1988.}\cr\noalign{\vskip2.truept}
LaL&{Landau, L.D., Lifshits, E.M.: {\it Statistical Physics}, Pergamon Press,
  1968.}\cr\noalign{\vskip2.truept}
LL&{Lieb, E., Lininger, W.: {\it Exact analysis of an interacting
  Bose gas. I. The general solution and the ground state}, Physical Review
  {\bf 130}, 1605--1624, 1963.}\cr\noalign{\vskip2.truept}
NN&{Nepomnyashchii, Yu.A., Nepomnyashchii, A.A.: {\it Infrared divergence in
  field theory of a Bose system with a condensate}, Soviet Phys. JETP {\bf
  48}, 493-501, 1978.}\cr\noalign{\vskip2.truept}
NO&{Negele, J.W., Orland, H.: {\it Quantum many-particle systems},
  Addison-Wesley, 1987.}\cr\noalign{\vskip2.truept}
P&{Popov, V.N.: {\it Functional integration in Quantum Field Theory and
  Statistical Mechanics}, Reidel, Dordricht, 1983.}\cr\noalign{\vskip2.truept}
PS&{Popov, V.N., Seredniakov, A.V.: {\it Low-frequency asymptotic form of
  the self-energy parts of a superfluid Bose system at $T=0$}, Soviet Phys.
  JETP {\bf 50}, 193--195, 1979.}\cr\noalign{\vskip2.truept}
}

\bye